\documentclass[english,structabstract]{aa}
\usepackage{mathptmx}
\usepackage{amsmath}
\usepackage[T1]{fontenc}
\usepackage[varg]{txfonts}
\usepackage{babel}
\usepackage{array}
\usepackage{textcomp}
\usepackage{multirow}
\usepackage{amsmath}
\usepackage{amssymb}
\usepackage{amsfonts}

\usepackage{graphicx}
\usepackage{natbib}
\bibpunct{(}{)}{;}{a}{}{,} 

\newcommand{\lyxmathsym}[1]{\ifmmode\begingroup\def\b@ld{bold}
  \text{\ifx\math@version\b@ld\bfseries\fi#1}\endgroup\else#1\fi}

\bibpunct{(}{)}{;}{a}{}{,} 
\makeatother
\usepackage{color}

\begin{document}

    
\title{Kinematics of the local disk from the RAVE survey and the Gaia first data release}

\subtitle{}

\author{Annie C. Robin\inst{1} \and Olivier Bienaym\'e\inst{2} \and Jos\'e G. Fern\'andez-Trincado\inst{1} \and C\'eline Reyl\'e
}

\institute{Institut Utinam, CNRS UMR6213, Univ. Bourgogne Franche-Comt\'e, OSU THETA , Observatoire de Besan\c{c}on, BP 1615, 25010 Besan\c{c}on Cedex, France  \\\email{annie.robin@obs-besancon.fr}
\and
Observatoire Astronomique de Strasbourg,  Universit\'e de Strasbourg, CNRS, 11 rue de l'Universit\'e, F-67000 Strasbourg, France
}

\offprints{A.C. Robin}

\date{Received ...; Accepted...}

\date{}

\abstract{}{We attempt to constrain the kinematics of the thin and thick disks using the Besan\c{c}on population synthesis model together with RAVE DR4 and Gaia first data release (TGAS).}{ The RAVE fields were simulated by applying a detailed target selection function and the kinematics was computed using velocity ellipsoids depending on age in order to study the secular evolution. We
accounted for the asymmetric drift computed from fitting a St\"ackel potential to orbits. Model parameters such as velocity dispersions, mean motions, and velocity gradients were adjusted using an ABC-MCMC method. We made use of the metallicity to enhance the separation between thin and thick disks.} 
{We show that this model is able to reproduce the kinematics of the local disks in great detail. The disk follows the expected secular evolution, in very good agreement with previous studies of the thin disk. The new asymmetric drift formula, fitted to our previously described St\"ackel potential, fairly well reproduces the velocity distribution in a wide solar neighborhood. The $U$ and $W$ components of the solar motion determined with this method agree well with previous studies. However, we find a smaller $V$ component than previously thought, essentially because we
include the variation of the asymmetric drift with distance to the plane. The thick disk is represented by a long period of formation (at least 2 Gyr), during which, as we show, the mean velocity increases with time while the scale height and scale length decrease, very consistently with a collapse phase with conservation of angular momentum. }
{This new Galactic dynamical model is able to reproduce the observed velocities in a wide solar neighborhood at the quality level
of the TGAS-RAVE sample, allowing us to constrain the thin and thick disk dynamical evolution, as well as determining the solar motion. }
{}

\keywords{Galaxy:evolution, Galaxy:dynamics, Galaxy:disk, Galaxy:kinematics, Galaxy: formation, Galaxy: stellar content}

\maketitle

{}

\newcommand{\Msun}{$M_\odot~$}
\newcommand{\Sun}{$_\odot~$}
\renewcommand{\deg}{$^{\circ}$}
\newcommand{\Ro}{$R_{\odot}~$}
\newcommand{\Rgal}{$R_{\mathrm{gal}}$}
\newcommand{\zgal}{$z_{\mathrm{gal}}$}
\newcommand{\feh}{[\text{Fe/H}] }

\section{Introduction}

The kinematics of the local disk has been a popular subject of debate for several decades. Several aspects have been considered, such as how to distinguish the solar motion from the local standard of rest, how fast the secular evolution is, whether the thin disk is separated from the thick disk in velocity space, the local importance of the dark matter halo and its imprint on the stellar kinematics, where the vertex deviation comes from and how it relates to the spiral structure, the local effect of resonances of the bar and spiral arms, and the effect of the radial migration of the populations over time.

Recent large-scale spectroscopic surveys now provide the opportunity to understand these problems better. Numerous direct analyses of such surveys (GCS \citep{GCS}, RAVE \citep{Kordopatis2013AJ....146..134K,Kunder2017AJ....153...75K}, APOGEE \citep{Eisenstein2011}, Gaia-ESO \citep{GaiaESO}, among the largest ones) are giving outstanding contributions to theses studies, thanks to the large statistics that are possible with several hundred thousands of measurements homogeneous on the sky, in opposition to earlier studies that worked with samples of a few hundred stars at most. 

Inverse methods have been used to deduce the solar motion, secular evolution, based on distance estimates to stars (generally photometric distances, sometime spectrophotometric distances, and rarely parallaxes). The methods also use age estimates in order to deduce the secular evolution of the stellar populations. However, these estimates (ages and distances) are generally strongly biased because the errors on ages and distances have strongly dissymmetric distributions that are far from Gaussian. Moreover, the sample selection functions also introduce their own bias.

We here consider a complementary approach to analyze the kinematics of a local sample using the population synthesis approach \citep{Creze1983nssl.conf..391C}. The method allows us to simulate the survey selection function and to avoid the use of photometric distances and ages by using instead the distribution of stars in the space of observables. The method is based on simple realistic assumptions for different populations and is constrained by the Boltzmann equation \citep{Bienayme1987a}. 

The method makes use of the latest version of the Besan\c{c}on galaxy model, hereafter BGM \citep{Robin2014,Czekaj2014} and is applied on the
 RAVE survey (DR4, \cite{Kordopatis2013AJ....146..134K}) and proper motions from the TGAS part of the Gaia DR1 \citep{Brown2016A&A...595A...2G} in order to constrain the kinematics of the different populations. We apply an approximate Bayesian computation Markov chain Monte Carlo (ABC-MCMC, \cite{marin:pudlo:robert:ryder:2011}) scheme to adjust the model parameters in the space of observables of the survey data. 

Section 2 describes the data sets and the selections applied on both the data and the simulations. The kinematical model parameters of the thin and thick disks are given in Section 3. The ABC-MCMC method is described in Section 4, while the results are reported and discussed in Section 5. Section 6 summarizes our main conclusions.

\section{Data set and selection function}

\subsection{Sample selection}

We make use of the RAVE data release DR4 \citep{Kordopatis2013AJ....146..134K}, which contains radial velocities, astrophysics parameters, and abundances for more than 400,000 stars in a wide solar neighborhood (up to 2 kpc from the Sun). It also contains proper motions for a large number of these stars. However, since the publication of the TGAS catalog of proper motions coming from the Gaia mission first release, we instead used these space-based proper motions, which are much more accurate and free of large systematics \citep{Arenou2017A&A...599A..50A}. The accuracy of the radial velocities is on the order of 2 km/s, while the proper motions are better than 1 mas/yr.
These data sets provide very accurate information on the 3D velocities of stars. 

The RAVE survey covers nearly the entire southern hemisphere, although in each field only a subset of available stars are measured, which is randomly selected in order to compose an unbiased sample. The stars are selected in bins of apparent $I$ magnitude that
well represents all types of stars in each magnitude bin. We limited the analysis to the range $9 < I <12,$ where the astrophysical parameters are accurate enough. Not all stars with velocities have measured metallicities. In particular at $I>11.5,$ the proportion of stars with reliable metallicities drops to about 40\%, while it is about 60\% at brighter magnitudes.
In order to use the metallicities and effective temperatures to better distinguish the populations, we used the RAVE sample of stars with metallicities [M/H]$_K$ available and temperatures between 3800 K and 8000 K. We also avoided using low-latitude fields ($|b|<25$\degr) because these regions have a complex target selection because of extinction, which is more difficult to reproduce correctly in the simulated sample. This limits the sample to 294,206 stars.

At the time of the submission of this publication, the RAVE DR5 was recently available with a preliminary version of the associated paper accessible on archiv.org, but this version was not yet accepted. For this reason, we maintained our analysis based on the RAVE DR4 data. In addition, our model-observations comparison is based half on proper motion and magnitudes, {taken from
other sources than RAVE}, while most RAVE radial velocities remain identical in DR4 and DR5.
In future works, we will base our analysis of the Galactic properties on the BGM and the DR5 since they present significant improvements such as 30,000 new stars, more accurate [M/H] and $T_{\rm eff}$ with bias corrected for the extreme values of [M/H] and $T_{\rm eff}$.

We can estimate the impact of using the DR4 instead of the DR5 release for the present analysis. DR5 data are improved thanks to a recalibration of the temperature, gravity, and metallicity. Figures 4, { 5, and 6} \citep{Kunder2017AJ....153...75K} allow
us to compare the mean DR5 versus DR4 values. { The difference
in gravity is not negligible, but we do not use it in the present study.} The definition of our subsamples for the
analysis of the observed proper motions and radial velocities is based on cuts in I magnitudes (the same magnitudes in DR4 and DR5), a cut in temperature at 5300 K, and cuts in metallicity [M/H], split at -1.2, -0.8, -0.4, and 0.  { Our [M/H] and $T_{\rm eff}$ cut values are not modified with the new calibrations of \cite{Kunder2017AJ....153...75K}. Hence, using the DR5 instead
would not have introduced} systematic changes in the content of our subsamples. Furthermore, the proper motions are extracted from the TGAS catalog, and radial velocities are not modified,
with the exception of some mismatch corrections.

TGAS proper motions are used for each star selected in the RAVE survey. It would have been possible to use the other TGAS stars, but at the expense of not having any information on metallicity and effective temperature, which two values are used here to separate thin- and thick-disk populations and dwarfs from giants. {Moreover, we found that the TGAS sample cross-identified with RAVE drops in completeness below a magnitude of 10. Hence we limited the analysis to this magnitude for the proper motion histograms.}

\subsection{Simulations}

The simulations were made using the revised version of BGM \citep{Czekaj2014}, where the thin-disk population is modeled with a decreasing star formation rate, a revised initial mass function (IMF), new evolutionary tracks and atmosphere models, and including the simulation of binarity.
For the thick-disk and halo population, the simulations are based on 
 \cite{Robin2014}, where the thick-disk structural and age parameters have been constrained together with the halo from color-magnitude diagrams fitting to SDSS and 2MASS photometry. We adopt here the thick-disk model shape B (secant squared in \zgal) and a Hernquist halo with a core radius of 5.17~kpc and an axis ratio of 0.776. \cite{Haywood2013} proposed that the probable period of formation of the thick disk extended over 3-4 Gyr, from 9 to 13 Gyr. 
In \cite{Robin2014} we showed that a good description of the populations in the thick disk, to reproduce SDSS and 2MASS data, was obtained when it is simulated by an extended star formation period, which can be simplified in the sum of two episodes, one about 12 Gyr ago, and the second one 10 Gyr ago. However, there is no evidence of two separate episodes, but rather a continuity between these two isochrones.
The best model was obtained when the thick-disk older stars are located in a wider structure, and the younger stars in a smaller structure. The scales in the older phase were typically estimated to be 2.9 kpc and 0.8 kpc for the scale length and scale height, respectively, and in the younger phase they were estimated to
be 2.0 kpc and 0.33 kpc (assuming exponential radially and sech$^2$ vertically). { Moreover, the ratio in local density of these two phases was found to be 0.15. Hence the young thick disk is the dominant component, while the old thick disk can be considered as marginal, although it is well detected in large surveys. It might coincide with the metal-weak thick disk identified previously by \cite{Norris1985ApJS...58..463N}. From the point of view of the kinematics,} if this scenario is correct, we should be able to estimate a difference in rotation between these two phases, that is to say, the older wider phase should rotate more slowly than the younger phase. 

We performed simulations in every RAVE field in the same photometric system and randomly selected the same number of stars as observed by RAVE in each $I$ -magnitude bin defined by RAVE, that is, 9-9.5, 9.5-10, 10-10.5, 10.5-10.8, 10.8-11.3, 11.3-11.7, and 11.7-12.

\begin{figure}[htb]
\begin{center}
\includegraphics[width=9cm]{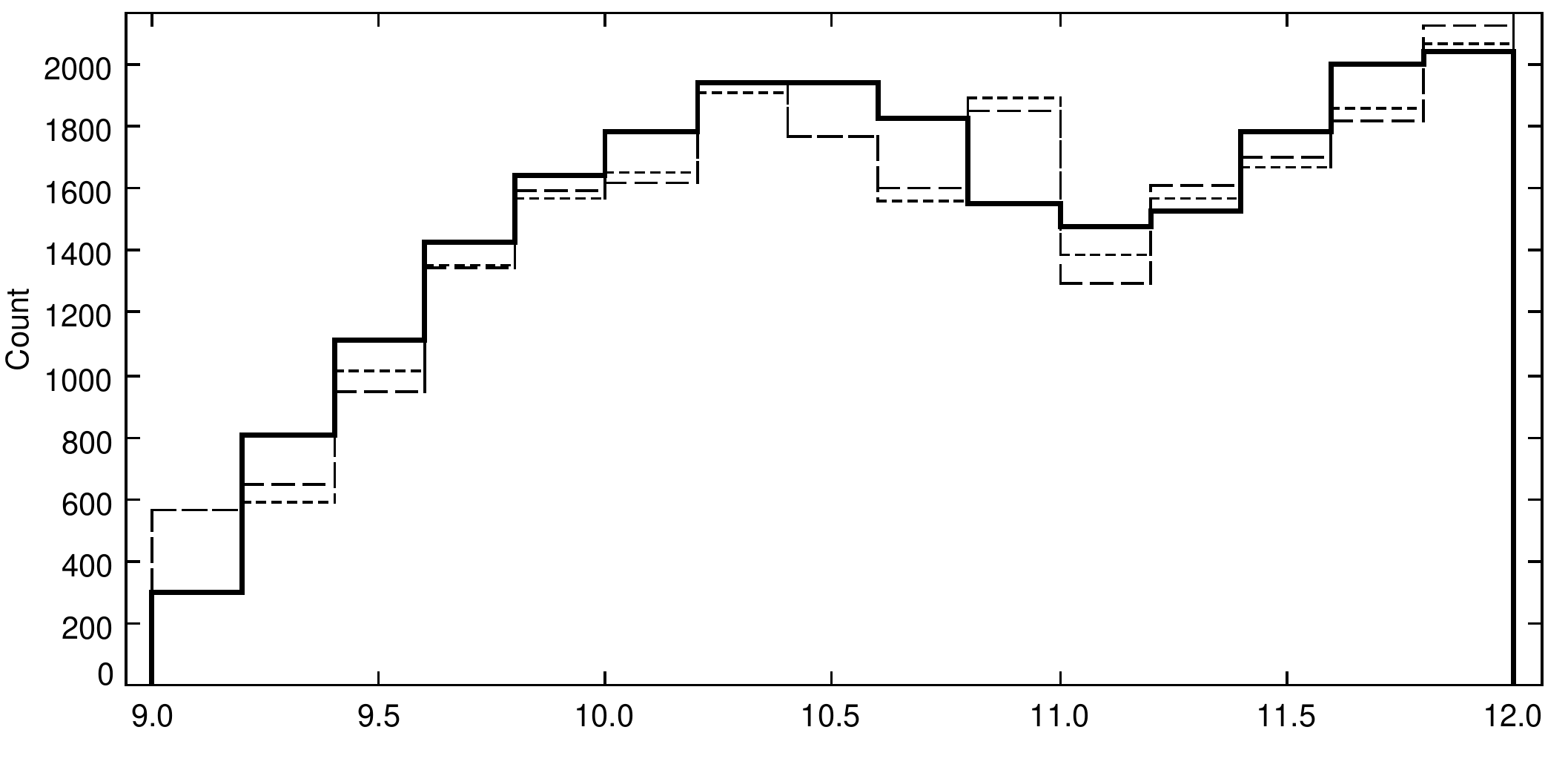}
\caption{Distribution in $I$ magnitude of the observed sample (solid line) of stars with reliable radial velocities and metallicities, and simulated sample (dashed  and dotted line for two independent simulated samples) for the whole RAVE survey.}
\label{fig:completeness}
\end{center}
\end{figure}

\begin{figure*}[htb]
\begin{center}
\includegraphics[width=9cm]{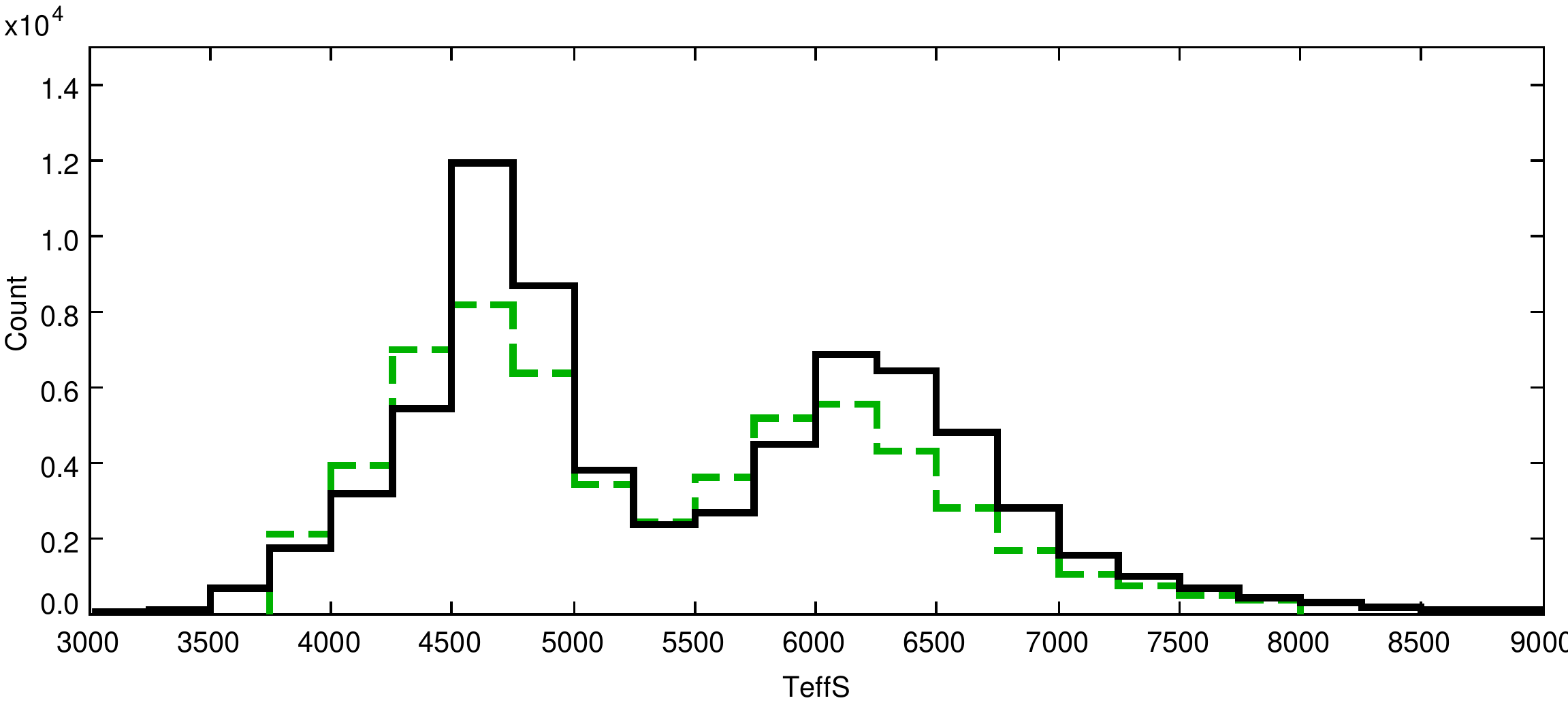}
\includegraphics[width=9cm]{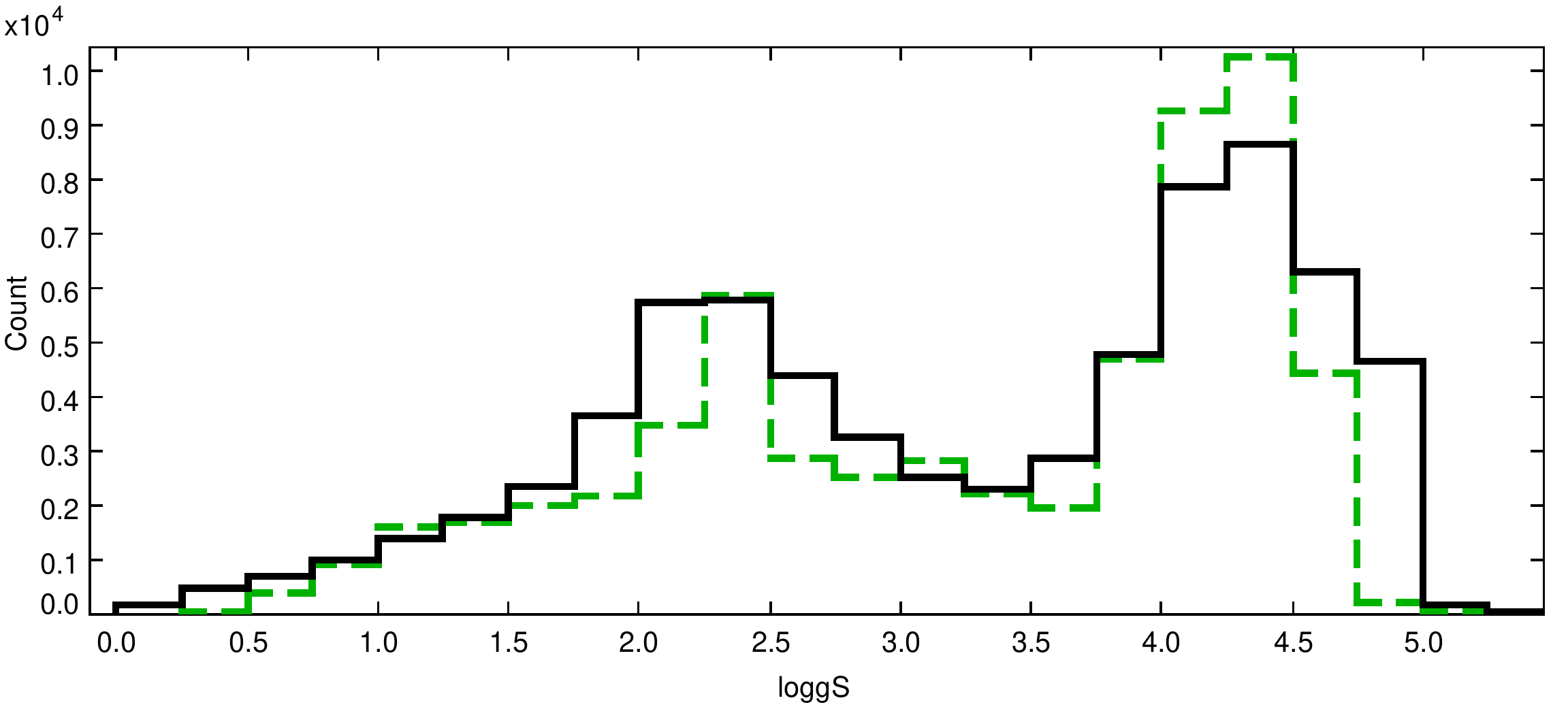}
\caption{Distribution in effective temperature (left panel) and gravity (right panel) of the observed sample (solid line) of stars with reliable radial velocities and metallicities, and simulated sample (dashed line) for the selection $|b|>$20\deg and $I<12$.}
\label{fig:Teff}
\end{center}
\end{figure*}

Figure~\ref{fig:completeness} shows the comparison of the $I$-magnitude distribution of the selected samples in observations and simulations. The distribution in $I$ in the simulation is close to the distribution of the real data. In order to check the distribution in astrophysical parameters of the whole sample, Figure~\ref{fig:Teff} presents the histogram of the distribution in effective temperature and the gravity of the simulated and observed sample.

In some cases the number of stars with reliable radial velocities and metallicities in RAVE data is on the order of 50 to 150 per field. This implies a significant Poisson noise when performing comparisons in metallicity bins. Therefore, we used simulations that have five times the number of observed stars  to decrease the Poisson noise, and we performed eight independent runs starting from random initial values in order to avoid to be stacked in local maxima of the likelihood.

 \section{Basis of the kinematical model}
 
 The kinematics of the stars is computed from simple assumptions of the relations between ages and velocity ellipsoids for the thin disk and {\it ad hoc} empirical values for the thick disk and the halo, as described below. The kinematics in the bar is more complex. In another study we use the bar potential to derive the velocity distributions from test-particle simulations (Fern\'andez-Trincado et al. in prep). For the purpose of the present work, the bar kinematics is not relevant since the bar population does not reach the RAVE sample in significant numbers. Although the bar potential can perturb the kinematics of the disk stars in the solar neighborhood, we here provisionally considered  a local axisymmetric potential. We adopted the usual reference system, with $U$ toward the Galactic center, $V$ toward rotation, and $W$ toward the North Galactic Pole.

\subsection{Thin disk}

The thin-disk kinematics is based on the velocity ellipsoids with dispersions varying with age, following the study of \cite{Bienayme1987a}.  
Table~\ref{age-sig0} gives the assumed age-velocity dispersion relation used in the previous Besan\c{c}on Galaxy Model \citep{Robin2003} following the determination of \cite{Gomez1997}  from the Hipparcos sample. The velocity dispersions increase with age, as expected for a secular evolution from a circular disk. 
\cite{Holmberg2009} proposed velocity dispersions slightly higher than \cite{Gomez1997}. 
RAVE and TGAS data provide a good opportunity to derive a more accurate age-velocity dispersion relation. We here considered that the velocity dispersion increases with age following three different formulae: 1) a third-order polynomial (Eq. \ref{eq1}); 2) the square-root formula proposed by \cite{Wielen1977A&A....60..263W} (Eq. \ref{eq2}); 3) the power-law formula taken from \cite{Aumer2016MNRAS.462.1697A} (Eq. \ref{eq3}).

\begin{equation} \label{eq1}
\sigma_W = A + B\times \tau + C \tau ^2
\end{equation}

\begin{equation}\label{eq2}
\sigma_W = \sqrt{\alpha + \gamma\times \tau }
, \end{equation}

\begin{equation}\label{eq3}
\sigma_W = k \times \tau ^\beta
,\end{equation}

\noindent where $\tau$ is the age in Gyr.

Then the velocity dispersion in $U$ and $V$ are defined relative to $\sigma_W$ by the ratios $\sigma_U/\sigma_V$ and $\sigma_U/\sigma_W$ , which are free parameters and assumed to be independent of time.

\begin{table}

\caption{Velocity ellipsoid as a function of age for the seven components of the thin disk in the standard Besan\c{c}on galaxy model.}
\label{age-sig0}
\begin{center}
\begin{tabular}{lllll}
\hline
 Subcomponent &Age range & $\sigma_{U}$ & $\sigma_{V}$  & $\sigma_{W}$  \\
  &Gyr & km/s & km/s & km/s \\
\hline
1 &0.-0.15 &  16.7   &  10.8    &  6.0\\
 2 &0.15- 1. & 19.8   &  12.8   &   8.0\\
3 &1.-2. &  27.2  &   17.6   &  10.0\\
 4 &2. - 3. & 30.2 &    19.5  &   13.2\\
 5 &3.- 5. & 36.7  &   23.7   &  15.8\\
 6 &5.- 7. & 43.1  &   27.8   &  17.4\\
 7 &7. -10. & 43.1  &   27.8   &  17.5\\
\hline
\end{tabular}
\end{center}
\end{table}
 
 In the fitting process we also considered the vertex deviation, allowing for two different angles $VD_a$ and $VD_b$ for stars with ages younger or older than 1 Gyr, respectively.
 
 \subsection{Thick disk}
 
The thick-disk velocity ellipsoid suffers from uncertainties that are mainly due to the way that this population is selected in different data sets. The relative continuity (or lack of clear separation) of the two disk populations, when no elemental abundances are available, has been the source of misunderstanding and apparent inconsistency between various analyses. With the help of the
alpha-element abundance ratio it can be easier to analyze, since it has been shown in local samples \citep{Adibekyan2013} and more distant samples \citep{Hayden2014}, among others, that the thick disk can be better separated from the thin disk using the $[\alpha/Fe]$ ratio. However, the abundances in RAVE DR4 are not accurate enough to clearly distinguish the thick disk from the thin-disk sequence in the $[\alpha/Fe]$ versus [Fe/H] plane. After several tests, we decided to use only [Fe/H] to separate the populations in the present analysis.
 
 According to \cite{Robin2014}, the scale length and scale height change from the beginning (12 Gyr ago) to the end of the phase (10 Gyr ago) because of contraction. Hence we expect that the velocity ellipsoid and rotation velocity show a similar behavior. 
 
 For the present analysis we chose to keep the velocity dispersions of the thick disk as free parameters together with its rotation velocity, but we assumed that these parameters are different for the two thick-disk episodes, mimicking a time evolution.
 The fourth-order dependency on time of the velocity dispersion ellipsoids takes this into account.
 
 The other populations (halo and bar) are very marginal in the
RAVE survey and will not {change the result of} the analysis.
 
 \subsection{Rotation curve and asymmetric drift}
 
\begin{figure}
\begin{center}
\includegraphics[width=9cm,angle=0]{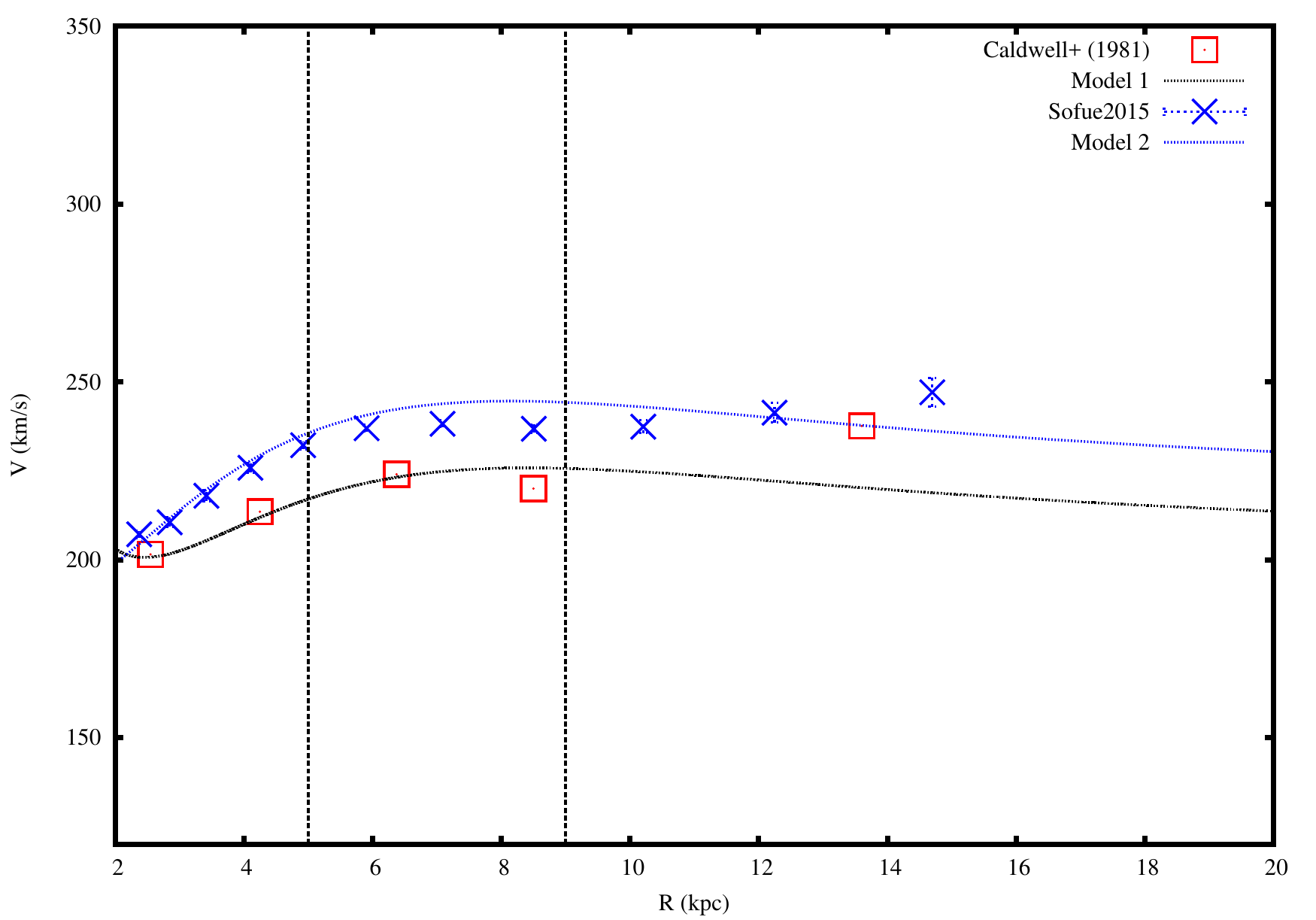}
\caption{Rotation curve of the mass model, compared with \cite{Caldwell1981ApJ...251...61C} and \cite{Sofue2015PASJ...67...75S}
. The interval between the two dashed lines indicates the range of Galactocentric distances implied in the present study.}
\label{rotation_curve}
\end{center}
\end{figure}

 In order to simulate the kinematics of stars at larger distances from the Sun, we used the rotation curve produced by the mass model. To compute the radial force, we summed the different mass components (stellar populations, interstellar matter, and dark matter halo) and derived the circular velocity as a function of Galactocentric radius. In this process described in \cite{Bienayme1987a} , we used observational data either from \cite{Caldwell1981ApJ...251...61C} or from \cite{Sofue2015PASJ...67...75S} to constrain the dark matter halo distribution and the thin-disk ellipsoid axis ratio. The resulting rotation curves are presented in Figure~\ref{rotation_curve}. { There is a significant difference between the two rotation curves, but at the solar position, their slopes are very similar. As we show below, they result in a similar fit to RAVE+TGAS data because these data mainly constrain the velocity dispersions, the slope of the rotation curve, and the asymmetric drifts, but set only weak constraints on the amplitude of the rotation curve itself.} 
 
The asymmetric drift was then computed to take the change of the circular velocity as a function of age and distance from the plane with regard to the rotation curve into account. It depends on the velocity dispersion ratio, on density and kinematic gradients, and on the difference of the radial force as a function of \Rgal~and \zgal. 

\begin{figure*}
\includegraphics[width=6cm,angle=270]{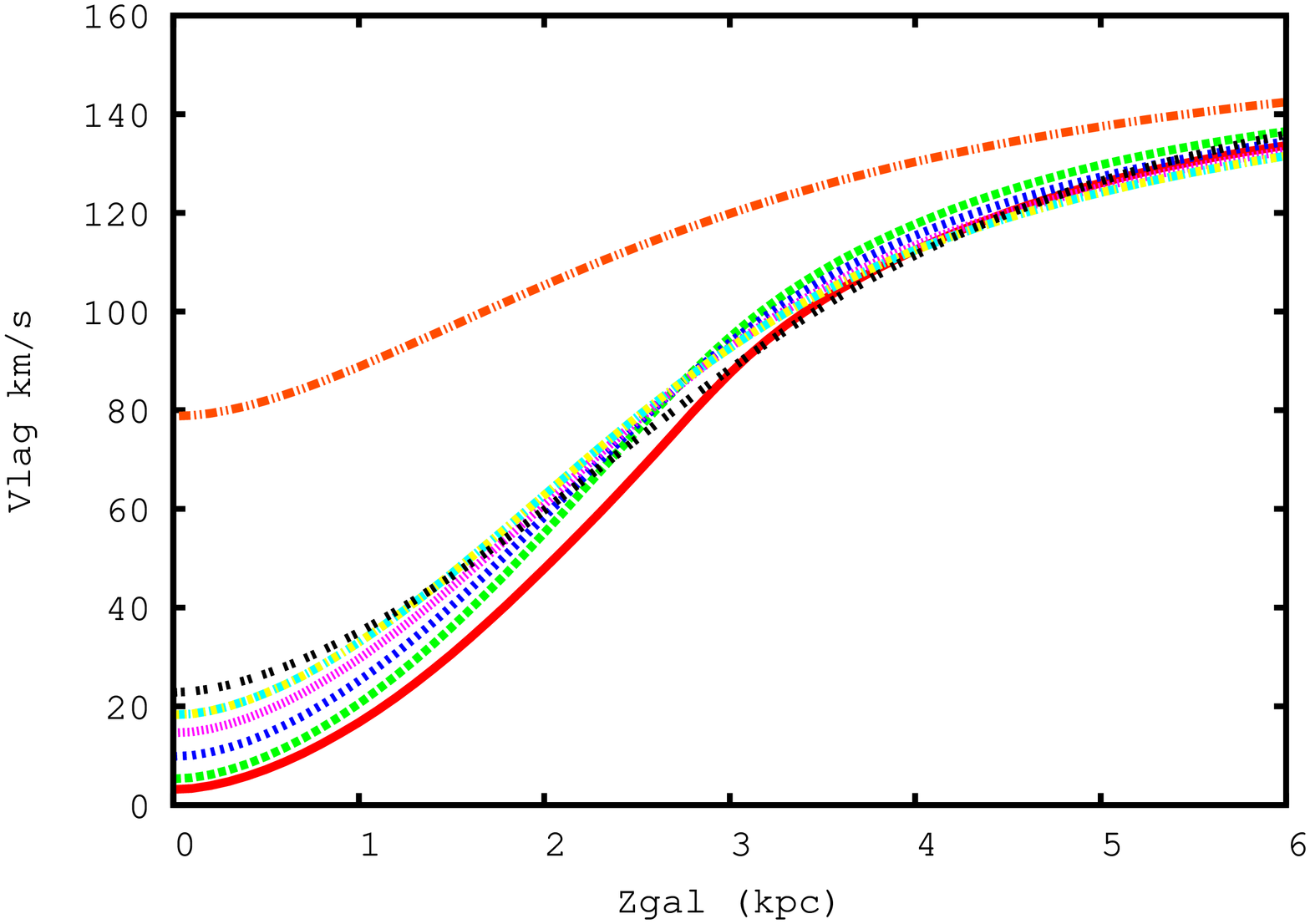}\includegraphics[width=6cm,angle=270]{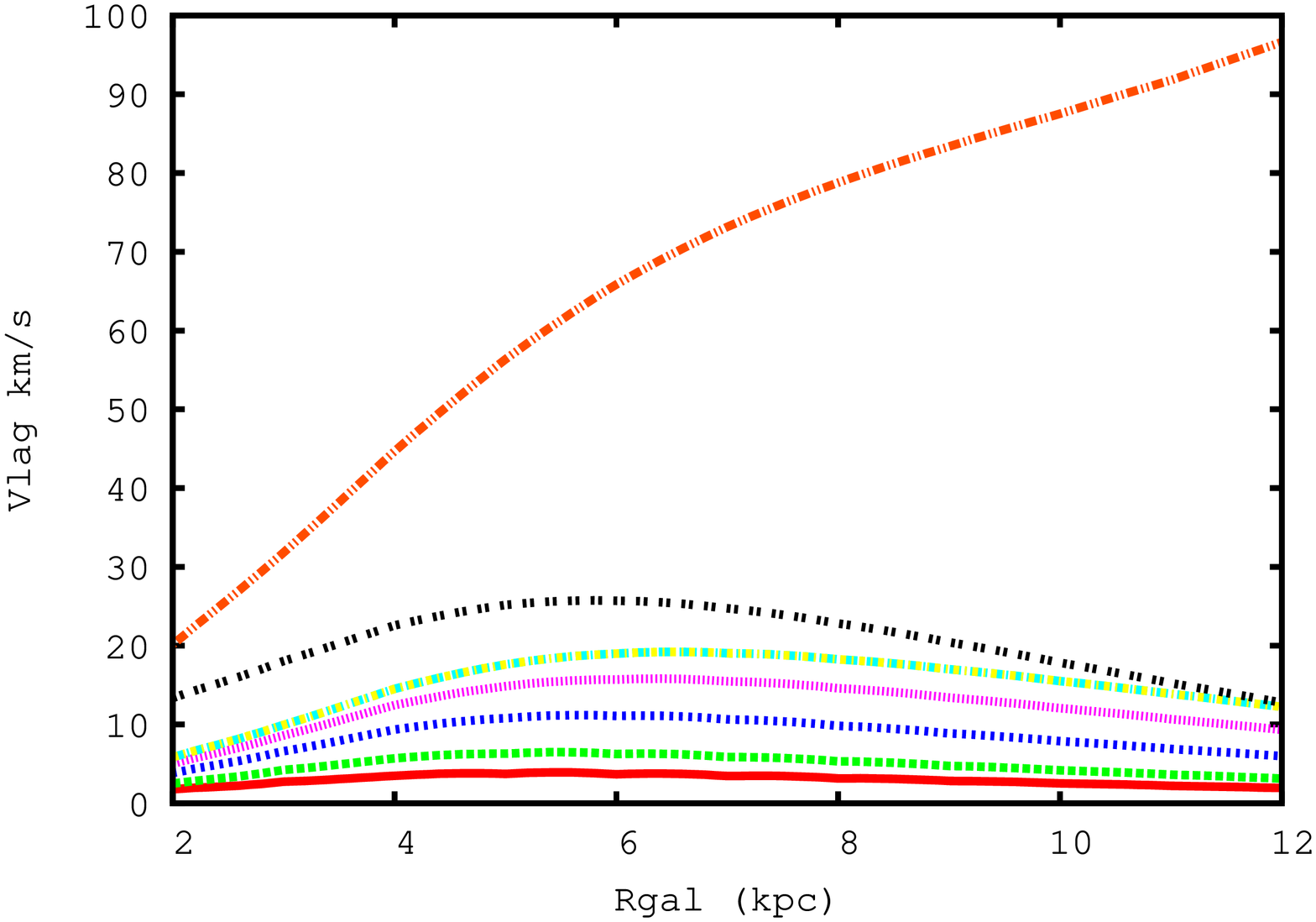}
\caption{Asymmetric drift computed from the St\"ackel approximation of the BGM for subcomponents 2 to 7 (thin disk, with increasing
ages plotted in solid red, long dashed green, short dashed blue, dotted magenta, dashed yellow and cyan), and for the young (dot-dashed black) and old (dot-dot-dashed red) thick disks. Left panel: as a function of \Rgal for \zgal=0; right panel: as a function of \zgal~for $R=R_\odot$.
}
\label{vca}
\end{figure*}

In the past we used the simplified formula of the asymmetric drift proposed by \cite{Binney2008}, which is valid in the Galactic plane. In order to have a more consistent expression of the asymmetric drift as a function of distance from the plane, we made use of the gravitational potential inferred by the mass distribution of the BGM. 
 \cite{Bienayme2015A&A...581A.123B} proposed distribution functions based on a St\"ackel approximation of the BGM potential, for which it is possible to compute a third integral of motion. An approximate fit of the variations of the asymmetric drift with ($R_{\mathrm{gal}}$, \zgal) Galactocentric coordinates was computed and used in the kinematical modeling. This approximation is valid in the range $2~$kpc $< R_{\mathrm{gal}} < 16$~kpc and $-6~$kpc $< z_\mathrm{gal} < 6$kpc and { was shown to be a very good approximation, as the $K_z$ is reproduced at better than 1\%}. The resulting rotational lag of different thin- and thick-disk subcomponents are shown to strongly depend on \Rgal~and \zgal. These dependencies are presented in Fig. \ref{vca}.
 
 \subsection{Solar velocities}
 
 There have been a number of studies that tried to measure the peculiar velocities of the Sun. The $U$ and $W$ velocities are relatively well known and have been constrained at a level of 1-2 km/s. This is not the case for the $V$ velocity, which is uncertain because it is difficult to distinguish it from the mean circular motion of the LSR. Hence, while in the past it was admitted to be about 5-6 km/s (see, i.e., \cite{Dehnen1998MNRAS.298..387D}), more recent studies found much higher values. For example, \cite{Schonrich2010} found about 12 km/s  {from a subset of the Geneva-Copenhagen Survey (GCS)}, while \cite{Bovy2012ApJ...759..131B} {from the APOGEE first data release \citep{Ahn2014ApJS..211...17A}} found an even higher value of 26$\pm3$ km/s). These values are not independent of the tracer selection, {mostly K giants in the case of \cite{Bovy2012ApJ...759..131B} and  mostly dwarfs in the case of \cite{Schonrich2010}}. The values also significantly depend on the rotation curve that is assumed and on the distance between the Sun and the Galactic center. This is why in our analysis the solar motion is a free parameter that can be influenced by other parameters and by the way the asymmetric drift is modeled.
 
 \section{Setting up the MCMC}
 
 This study uses an ABC-MCMC code based on a Metropolis-Hasting sampling, as described in \cite{Robin2014}.
 Table~\ref{param} shows the set of model parameters to fit and their respective range. To estimate the goodness-of-fit for each model, we directly compared histograms of radial velocity and proper motions between the model and the data. The bin steps in radial velocity and proper motions are  5 km/s and 5 mas/yr, respectively. To improve the sensitivity of the analysis, however, we separated stars using their metallicity and temperature. The metallicity rather than $\alpha$ abundances was used because the accuracy in $\alpha$ in the RAVE survey does not allow us
to separate the thin disk well from the thick-disk sequences, and we expect that the metallicity is more accurate. Moreover, to separate the young thick disk from the old thick disk, the metallicity is more efficient than the $\alpha$ abundances, and
it also separate the metal-rich thin disk better from the normal thin disk.
  
 We made use of four metallicity bins, where lower metallicities are dominated by the thick disk and higher metallicities by the thin disk. The minimum metallicity is -1.2 and the width of each bin is 0.4 dex. To separate dwarfs from giants, we cut at a temperature of 5300 K, considering that cooler stars are mainly giants, while hotter stars are mainly dwarfs. However, we did not explicitly assume that they are dwarfs or giants in the analysis. We merely
considered these two bins and computed histograms of velocities and likelihoods for these two bins separately in both data and simulations to enhance the efficiency of the fit of the parameters depending on ages. For the proper motions we used projections on Galactic coordinates, $\mu_l^* = \cos(b) \times \mu_l$ and $\mu_b$, but for the South Galactic cap it is more interesting to project the proper motions parallel to the $U$ and $V$ vectors, $U$ pointing toward the Galactic center, and $V$ toward rotation. This facilitates the interpretation of the histogram comparisons
because it clearly shows the skewness of the $V$ distribution that is due to the asymmetric drift.
 
The likelihoods were computed separately for 11 regions of the sky, corresponding to different latitudes and quadrants in longitudes. The likelihood is based on the formula given in \cite{Bienayme1987a} for a Poisson statistics.  Then, to intercompare different models with a different number of free parameters, we computed the Bayesian information criterion (BIC) following \cite{Schwarz1978}, which penalizes models with more free parameters (Eq. \ref{BIC}),
\begin{equation}
\label{BIC}
BIC = -2.\times Lr + k \times ln(n),
  \end{equation}
  where $Lr$ is the likelihood, $k$ is the number of parameters, and $n$ the number of observations used in the likelihood computation.
  
 \begin{table}
\caption{Set of parameters and range used in the ABC-MCMC process. Units are km/s for velocities and radian for the vertex deviation. Thin-disk parameters A, B, and C are the coefficients of the
formula describing the evolution of $\sigma_W$ with time (see text). The vertex deviation is for ages younger ($VD_a$) and older ($VD_b$) than 1 Gyr, and the scale lengths are given in kpc. Velocities and velocity dispersions are all given in km/s.} 
\label{param}
\begin{center}
\begin{tabular}{llll}
\hline
 Component & Parameter & min & max \\
\hline
{\it Solar motion} \\
& $U$\Sun & 0. & 20.\\
& $V$\Sun & 0. & 30.\\
& $W$\Sun& 0. & 20.\\
{\it Vertex deviation}\\
& $VD_a$ & -1 & 1\\
& $VD_b$ & -1 & 1\\

\hline
{\it Thin disk}\\
&$A$     & 4&  60\\
&$B$               & 0&  60\\
&$C$               & -0.5&  0.5\\
&$\sigma_{V}/\sigma_{U}$ &  0.3&  1.\\
&$\sigma_{W}/\sigma_{U}$ &  0.3&  1.\\
&$h_{\sigma_{U}}$                  & 1 & 25.\\
&$h_{\sigma_{W}}$                  & 1 & 25.\\
\hline
{\it Thick disk}\\
&$\sigma_{U}$             &  25. & 80.\\
&$\sigma_{V}$             &    25. & 80\\
&$\sigma_{W}$             &    25. & 80\\
\hline
{\it Old thick disk}\\
&$\sigma_{U}$             &    25. & 80\\
&$\sigma_{V}$             &     25. & 80\\
&$\sigma_{W}$             &    25. & 80\\
\hline
\end{tabular}
\end{center}
\end{table}

At the end of the process, we considered the last third of each Markov chain, containing 200,000 iterations each,  for eight independent runs of the MCMC and computed the mean and dispersion for each fitted parameter.   

\section{Results}

The values of the fitted parameters are given in Tables~\ref{res-k1}, ~\ref{res-k2}, and ~\ref{res-k3} for the age-velocity dispersion as a fourth-order polynomial, square root formula, and power-law formula, respectively.
The velocity dispersions for the old thick disk are noticeably larger than the dispersion of the young thick disk, as expected from their respective scale heights.  This confirms the collapse with time during the thick-disk phase and the probable contraction from a larger thick disk with slower rotation in the past toward a smaller and more concentrated thick disk with faster rotation later on.

The solar velocities are also well constrained by this analysis, and we find mean motions in good agreement with previous studies for $U$ and $W$ velocities. For the case of the circular velocity, we obtained a lower value than has been found in many other studies. This is discussed in the next section. 

\begin{figure}
\begin{center}
\input{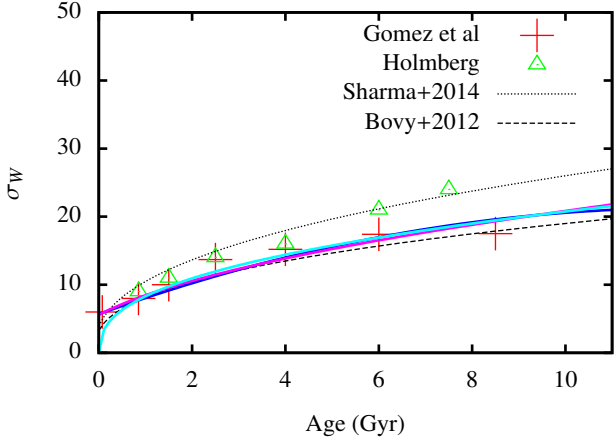}
\caption{Evolution of the {vertical} velocity dispersion of the thin disk with age. {The solid lines show the best-fit solutions for the three different formulae (fit 1: blue, fit 2: magenta, fit 3: cyan,} see text), while the symbols indicate the \cite{Gomez1997} values from Hipparcos (red plus) and \cite{Holmberg2009} (green triangles). {The black dotted line} is the relation from \cite{Sharma2014ApJ...793...51S}, while the {black dashed line is the relation }from \cite{Bovy2012ApJ...759..131B}.}
\label{disc-diffusion}
\end{center}
\end{figure}

The thin-disk diffusion with time is also constrained to be close to the values obtained by \cite{Gomez1997} from Hipparcos data.
Figure~\ref{disc-diffusion} shows the comparison between the results of our fits with the three assumed formulas and data from \cite{Holmberg2009} and \cite{Gomez1997}. The agreement is good for the young ages with both determinations but is closer to \cite{Gomez1997} for the older stars. We also overplot the velocity dispersion as a function of time obtained by \cite{Sharma2014ApJ...793...51S} and by \cite{Bovy2012ApJ...759..131B} for the further discussion.

 \begin{table}
\caption{Best values of fitted parameters obtained by the mean of the
  last third of eight independent chains and standard deviation assuming
  \cite{Caldwell1981ApJ...251...61C} and \cite{Sofue2015PASJ...67...75S} rotation curves. Units are km/s for velocities, pc for the scale lengths, and radians for the vertex deviation, which is given for stars younger than 1 Gyr ($VD_a$) and older than 1 Gyr ($VD_b$). A, B, and C are the coefficients of the polynomial describing the variation of $\sigma_W$ with age in Gyr (Eq. \ref{eq1}). The BIC is computed from Eq. \ref{BIC}.}
\label{res-k1}
\begin{center}
\begin{tabular}{lcc}
\hline
 Parameter & Caldwell & Sofue  \\
\hline
{\it Solar motion} \\
 $U$\Sun             & 12.75  $\pm$   1.26      &         11.88$\pm$  1.38\\                                                             
$V$\Sun             &  0.93  $\pm$   0.30       &                  0.91$\pm$  0.26\\     
$W$\Sun             &  7.10  $\pm$   0.16       &                  7.07$\pm$  0.16\\     
{\it Vertex deviation} \\                                                                   
$VD_a$              &   -0.0439  $\pm$     0.0375       &           -0.0618$\pm$    0.0218\\  
$VD_b$              &   -0.0144  $\pm$     0.0122       &           -0.0048$\pm$    0.0108\\  
\hline                                                                                      
{\it Thin disk }        \\                                                                           
A                       &  5.69  $\pm$   0.37   &                  5.69$\pm$  0.41\\     
B                       &  2.48  $\pm$   0.30   &                  2.33$\pm$  0.28\\     
C                       &   -0.0966  $\pm$     0.0404   &           -0.0774$\pm$    0.0362\\  
$\sigma_{V}/\sigma_{U}$ &  0.57  $\pm$   0.03   &                  0.58$\pm$  0.03\\     
$\sigma_{W}/\sigma_{U}$ &  0.46  $\pm$   0.03   &                  0.46$\pm$  0.02\\     
h$_{\sigma_{U}}$         &     13176.  $\pm$       6908.        &      9534.$\pm$      3982.\\      
h$_{\sigma_{W}}$         &     15919.  $\pm$       8609.        &     10414.$\pm$      6299.\\      
\hline                                                                                      
{\it Thick disk }       \\                                                                           
$\sigma_{U}$        & 40.02  $\pm$   1.74       &                 41.58$\pm$  1.51\\     
$\sigma_{V}$        & 31.86  $\pm$   1.55       &                 30.95$\pm$  1.50\\     
$\sigma_{W}$        & 27.89  $\pm$   1.26       &                 27.02$\pm$  1.00\\     
\hline                                                                                      
{\it Old thick disk}    \\                                                                  
$\sigma_{U}$        & 75.64  $\pm$   8.58       & 79.64$\pm$  7.96\\
$\sigma_{V}$        & 55.41  $\pm$   8.74       &                 57.55$\pm$  8.51\\     
$\sigma_{W}$        & 66.43  $\pm$   3.95       &                 62.15$\pm$  6.62\\     
\hline                                                                                      
$Lr$                    &     -5384.  $\pm$         38. &     -5378.$\pm$       155.\\      
$BIC $                    &    10861.  $\pm$      76.   &                     10851.$\pm$       161.\\
\hline

\end{tabular}
\end{center}
\end{table}

\begin{table}
\caption{Same as Table~\ref{res-k1}, but here $\alpha$ and $\gamma$ are the parameters of the square-root function of the velocity dispersion as a function of age in Gyr (Eq. \ref{eq2}).}
\label{res-k2}
\begin{center}
\begin{tabular}{lcc}
\hline
Parameter &Caldwell & Sofue   \\
\hline
{\it Solar motion} \\
 $U${\Sun}            & 12.92 $\pm$   1.14&                       12.79 $\pm$   0.85\\          
 $V$\Sun             &  0.92 $\pm$   0.21&                         0.96 $\pm$   0.25\\          
 $W$\Sun             &  7.08 $\pm$   0.14&                         7.12 $\pm$   0.17\\          
{\it Vertex deviation}\\                                                                   
$VD_a$              &   -0.0435 $\pm$     0.0148&                   -0.0458 $\pm$     0.0152\\  
$VD_b$              &   -0.0148 $\pm$     0.0087&                   -0.0133 $\pm$     0.0064\\  
\hline                                                                                     
{\it Thin disk}\\                                                                          
$\alpha$                       & 32.33 $\pm$   1.96&                     32.23 $\pm$   1.89\\      
$\gamma$                       & 40.41 $\pm$   2.74&                     42.55 $\pm$   2.38\\      
$\sigma_{V}/\sigma_{U}$ &  0.57 $\pm$   0.03&                      0.57$\pm$  0.03\\   
$\sigma_{W}/\sigma_{U}$ &  0.46 $\pm$   0.02&                      0.47 $\pm$   0.02\\          
h$_{\sigma_{U}}$         &      7493. $\pm$       7009.&               14776. $\pm$       7946.\\
h$_{\sigma_{W}}$         &      8578. $\pm$       5896.&               16432. $\pm$       9622.\\
\hline                                                                                     
{\it Thick disk}\\                                                                         
$\sigma_{U}$        & 39.05 $\pm$   3.04&                         41.23 $\pm$   1.56\\          
$\sigma_{V}$        & 32.31 $\pm$   1.44&                         32.90 $\pm$   1.87\\          
$\sigma_{W}$        & 28.61 $\pm$   1.27&                         26.92 $\pm$   1.12\\          
\hline                                                                                     
{\it Old thick disk }\\                                                                     
$\sigma_{U}$        & 80.03 $\pm$  10.92&                         81.31 $\pm$   8.73\\          
$\sigma_{V}$        & 57.35 $\pm$   7.29&                         57.98 $\pm$   7.46\\          
$\sigma_{W}$        & 61.89 $\pm$   6.09&                         59.03 $\pm$   8.25\\          
\hline                                                                                     
$Lr$                    &  -5395.  $\pm$     42.&                     -5417. $\pm$         29.\\
$BIC $                    &  10874  $\pm$  85.  &                     10916. $\pm$         58.\\

\end{tabular}
\end{center}
\end{table}

\begin{table}
\caption{Same as Table~\ref{res-k1} but where $k$ and $\beta$ are the parameters of the power law function of the velocity dispersion as a function of age in Gyr (Eq. \ref{eq3}).}
\label{res-k3}
\begin{center}
\begin{tabular}{lcc}
\hline
Parameter & Caldwell & Sofue    \\
\hline
{\it Solar motion} \\
$U{\sun}$              & 13.00 $\pm$   1.02 &                            13.12 $\pm$   1.47\\     
$V{\sun}$              &  0.94 $\pm$   0.23 &                            0.92 $\pm$   0.29\\      
$W{\sun}$              &  7.01 $\pm$   0.15 &                             7.03 $\pm$   0.18\\     
{\it Vertex deviation} \\                                                                         
$VD_a$               &   -0.0325 $\pm$     0.0131 &                        -0.0388 $\pm$     0.0126\\
$VD_b$                  &   -0.0173 $\pm$     0.0075 & -0.0167 $\pm$   0. 0111 \\
\hline                                                                                            
{\it Thin disk} \\                                                                                
$k$                       &  8.30 $\pm$   0.19 &                                      8.26 $\pm$   0.22\\ 
$\beta$                      &  0.40 $\pm$   0.02 &                                    0.40 $\pm$   0.02\\
$\sigma_{V}/\sigma_{U}$ &  0.54 $\pm$   0.02 &                       0.57 $\pm$   0.03\\           
$\sigma_{W}/\sigma_{U}$ &  0.46 $\pm$   0.02 &                       0.47 $\pm$   0.02\\           
h$_{\sigma_{U}}$         &     19430. $\pm$       3948. &             13009. $\pm$       8856.\\      
h$_{\sigma_{W}}$         &     11813. $\pm$       8010. &             10473. $\pm$       7058.\\      
\hline                                                                                            
{\it Thick disk} \\                                                                               
$\sigma_{U}$            & 40.36 $\pm$   2.00 &                    42.18 $\pm$   2.05\\                 
$\sigma_{V}$            & 32.85 $\pm$   1.44 &                    32.26 $\pm$   2.06\\                 
$\sigma_{W}$            & 27.03 $\pm$   1.20 &                   26.90 $\pm$   1.10\\                 
\hline                                                                                            
{\it Old thick disk } \\                                                                                  
$\sigma_{U}$            & 80.30 $\pm$  10.32 &                    75.87 $\pm$   5.52\\                 
$\sigma_{V}$            & 57.81 $\pm$   6.35 &                    53.93 $\pm$   5.87\\                 
$\sigma_{W}$            & 62.24 $\pm$   5.25 &                    66.22 $\pm$   3.23\\                 
\hline                                                                                            
$Lr$                    &     -5429 $\pm$         32 &               -5420. $\pm$         31.\\ 
$BIC $                    &   10942 $\pm$      65 &                    10923. $\pm$         62.\\
\hline
\end{tabular}
\end{center}
\end{table}

\begin{figure*}[htb]
\begin{center}
\includegraphics[width=8cm,angle=0]{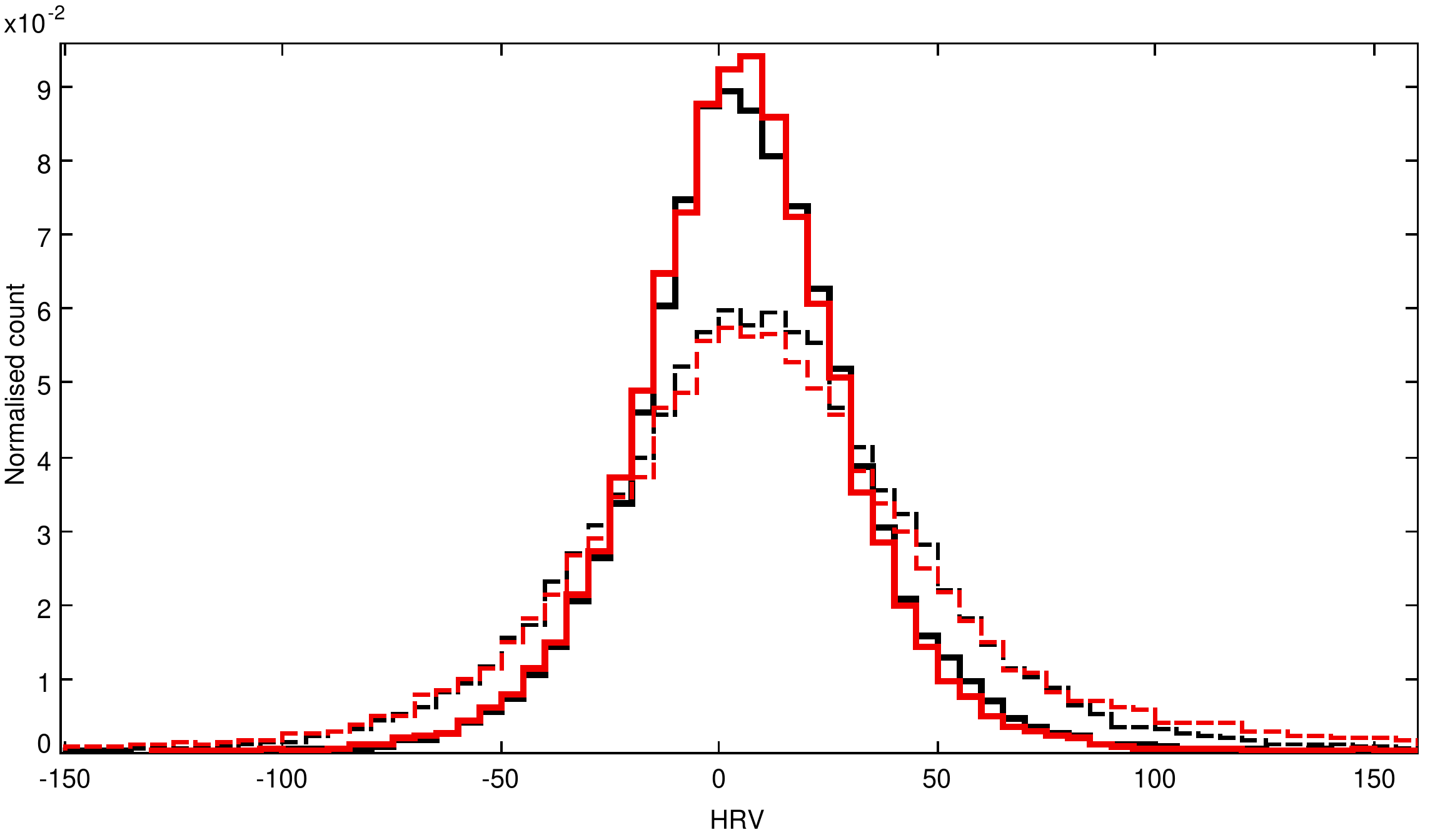}
\includegraphics[width=8cm,angle=0]{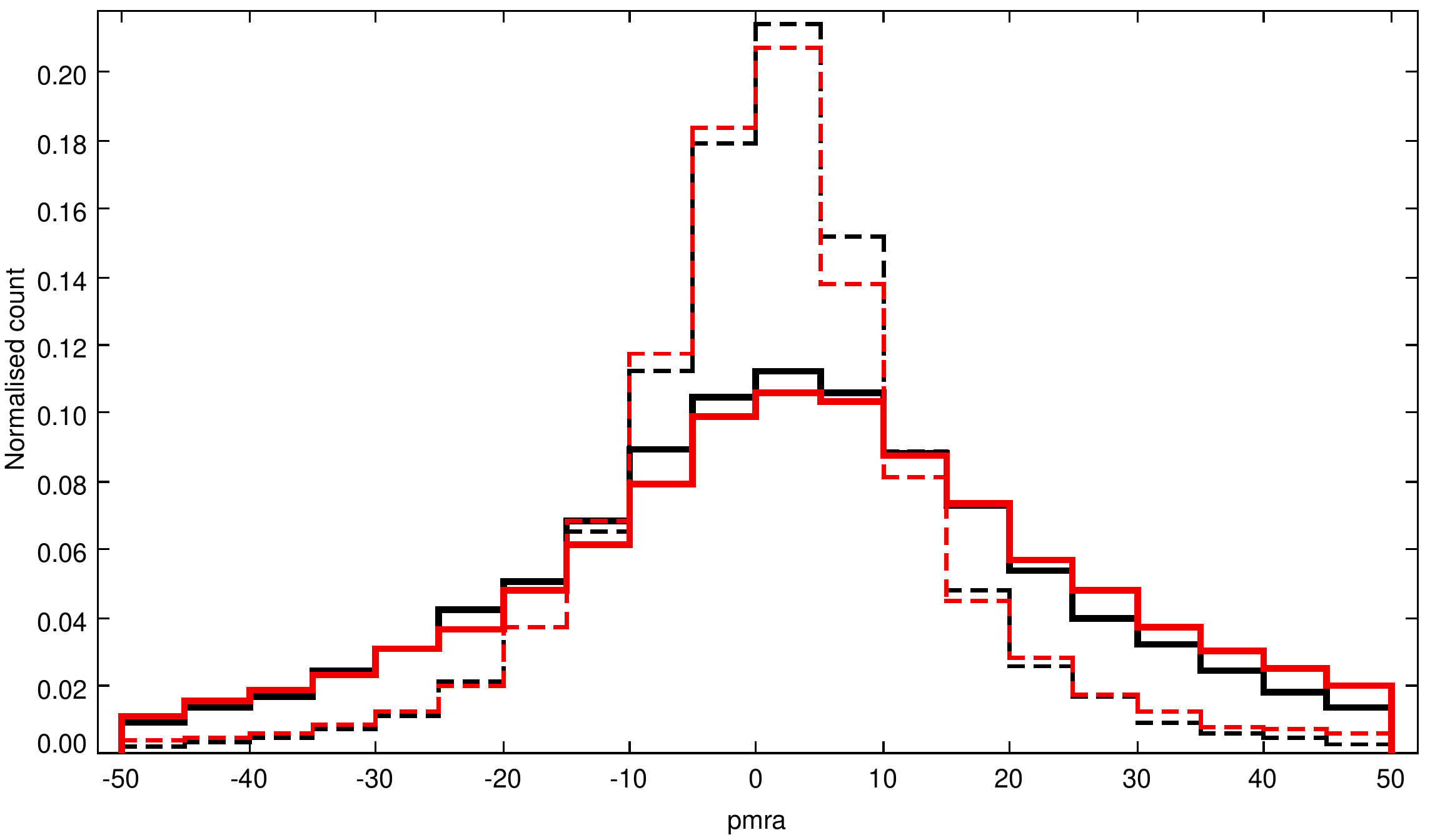}\includegraphics[width=8cm,angle=0]{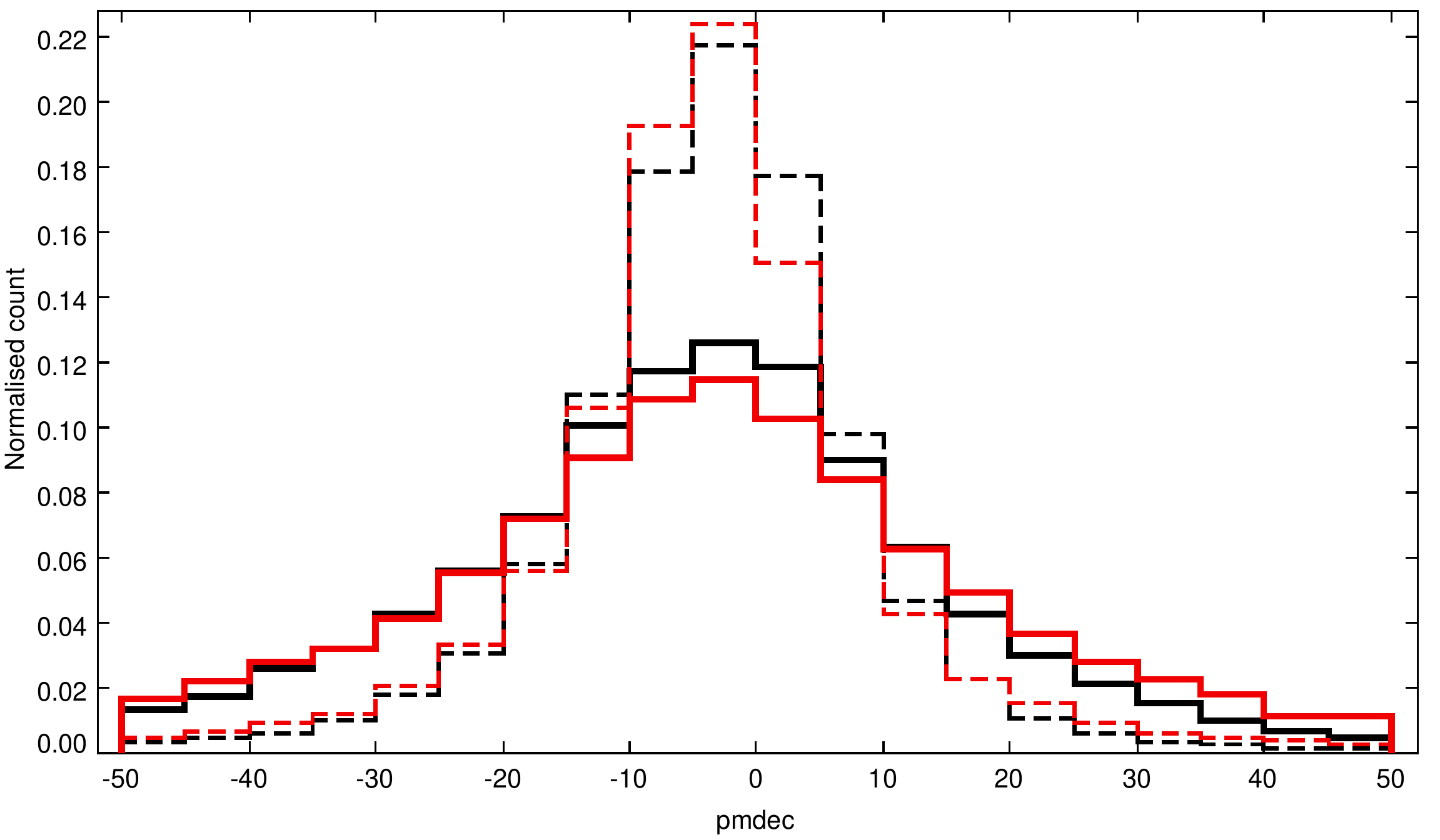}
\caption{Histograms of RAVE radial velocity distributions {(top panel) and TGAS proper motions (bottom panels: left: proper motion along the right ascension; right: along the declination)} for hot (solid lines) and cool (dashed lines) stars. Data are shown
as black lines, and the best-fit model is shown as red lines. }
\label{fig_cool_hot}
\end{center}
\end{figure*}

To visually evaluate the agreement between model and data, we show in Figure~\ref{fig_cool_hot} histograms for cool stars ($T_{\rm eff}$<5300K) and hot stars ($T_{\rm eff}$>5300K) of radial velocities and proper motions. We clearly see that cool stars present larger velocity dispersions (seen from radial velocity histograms), although in proper motions their dispersions are smaller because
of the distance effect in these mostly giant stars. 

\begin{figure*}[htb]
\begin{center}
\includegraphics[width=8cm,angle=0]{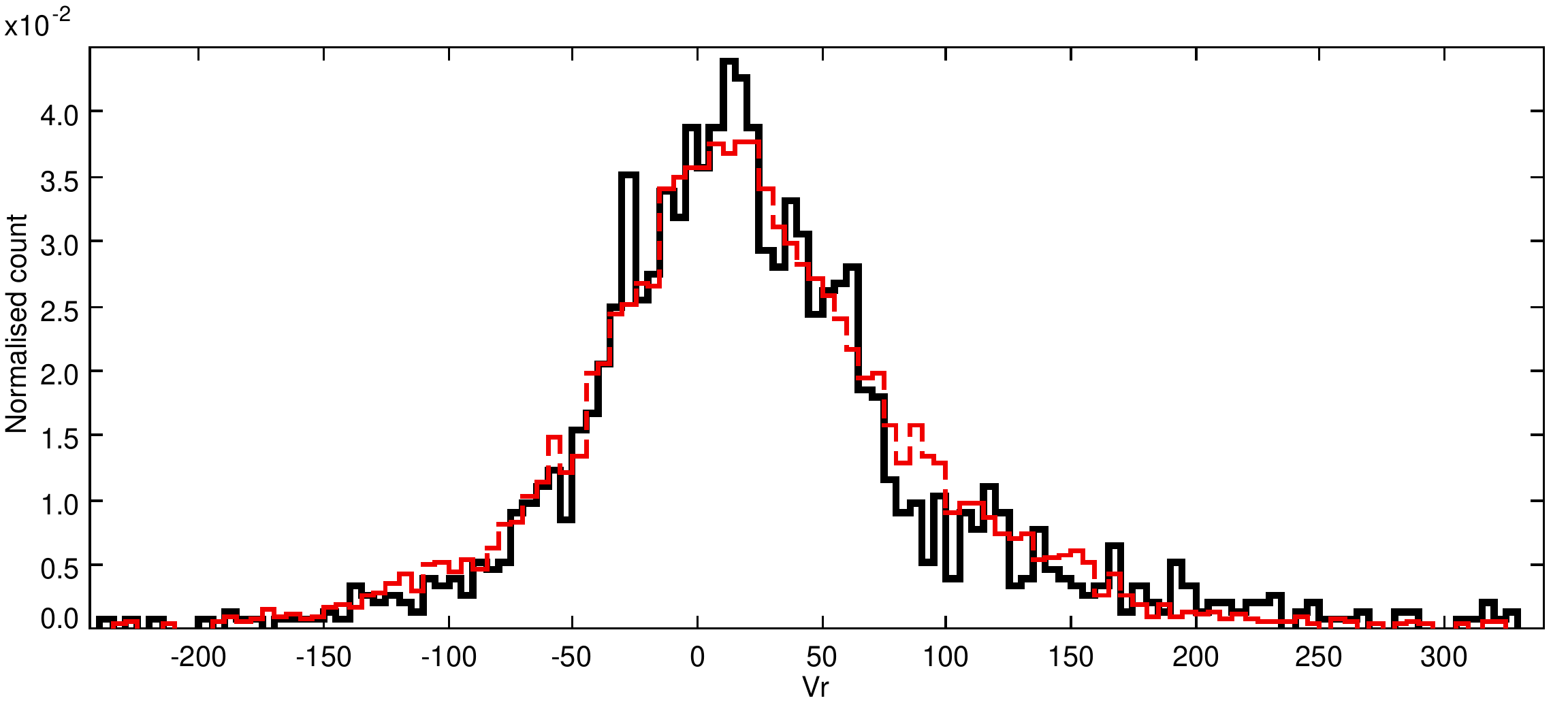}
\includegraphics[width=8cm,angle=0]{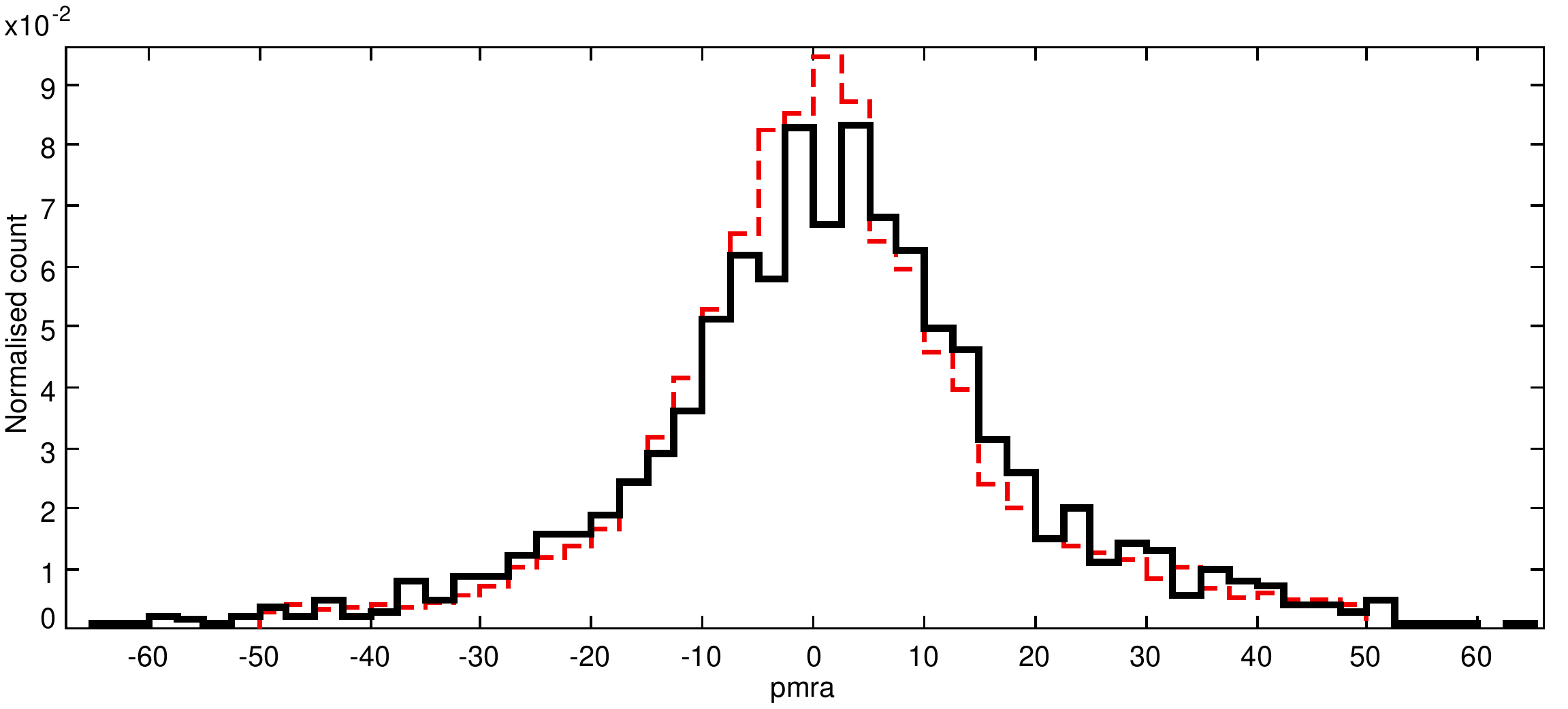}\includegraphics[width=8cm,angle=0]{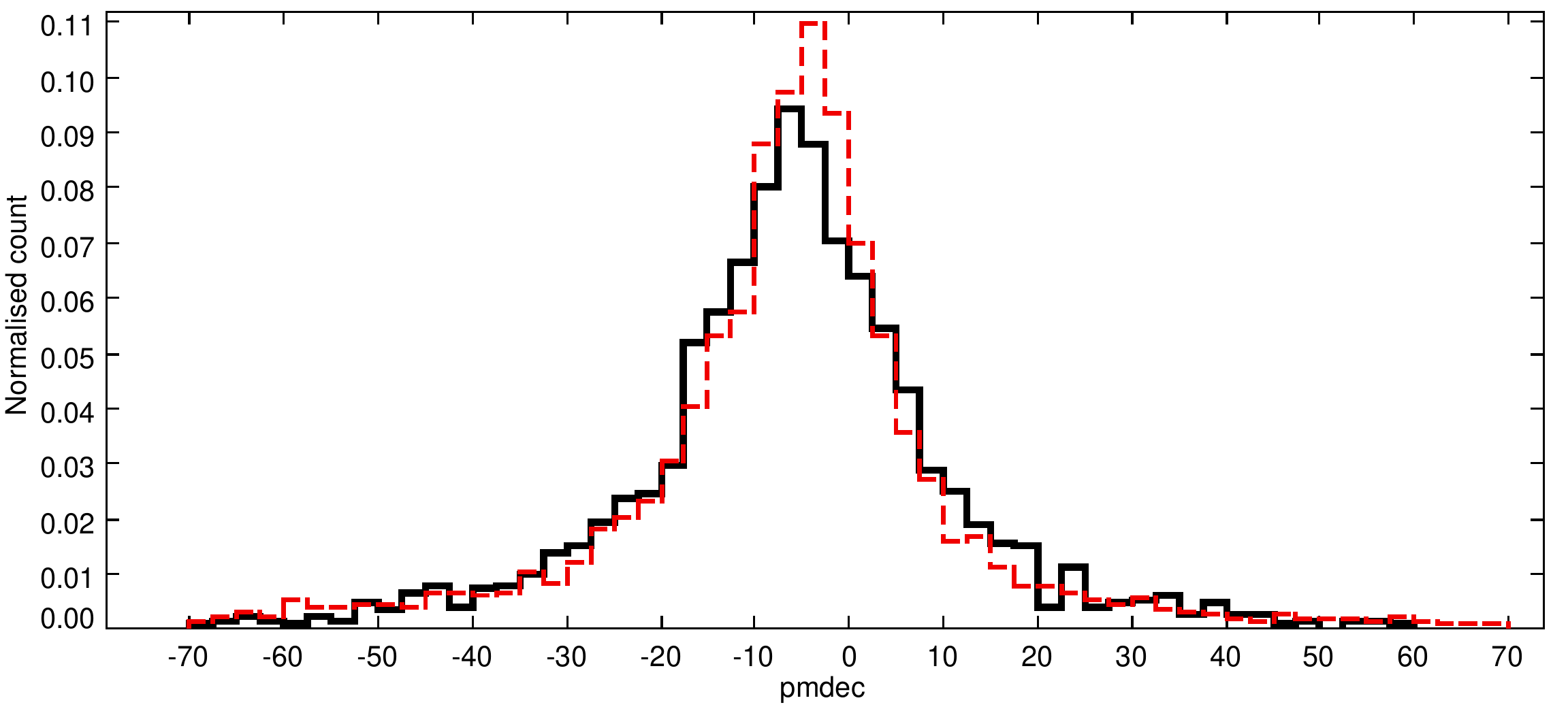}
\caption{Histograms of RAVE radial velocity distributions {(top panel) and TGAS proper motions (bottom panels: left: proper motion along the right ascension; right: along the declination)} for the metallicity bin -1.2 to -0.8 dex, dominated by the old thick disk. Data are shown as black solid lines, and the best-fit model
is shown as red dashed lines. }
\label{fig_feh-08}
\end{center}
\end{figure*}

\begin{figure*}[htb]
\begin{center}
\includegraphics[width=8cm,angle=0]{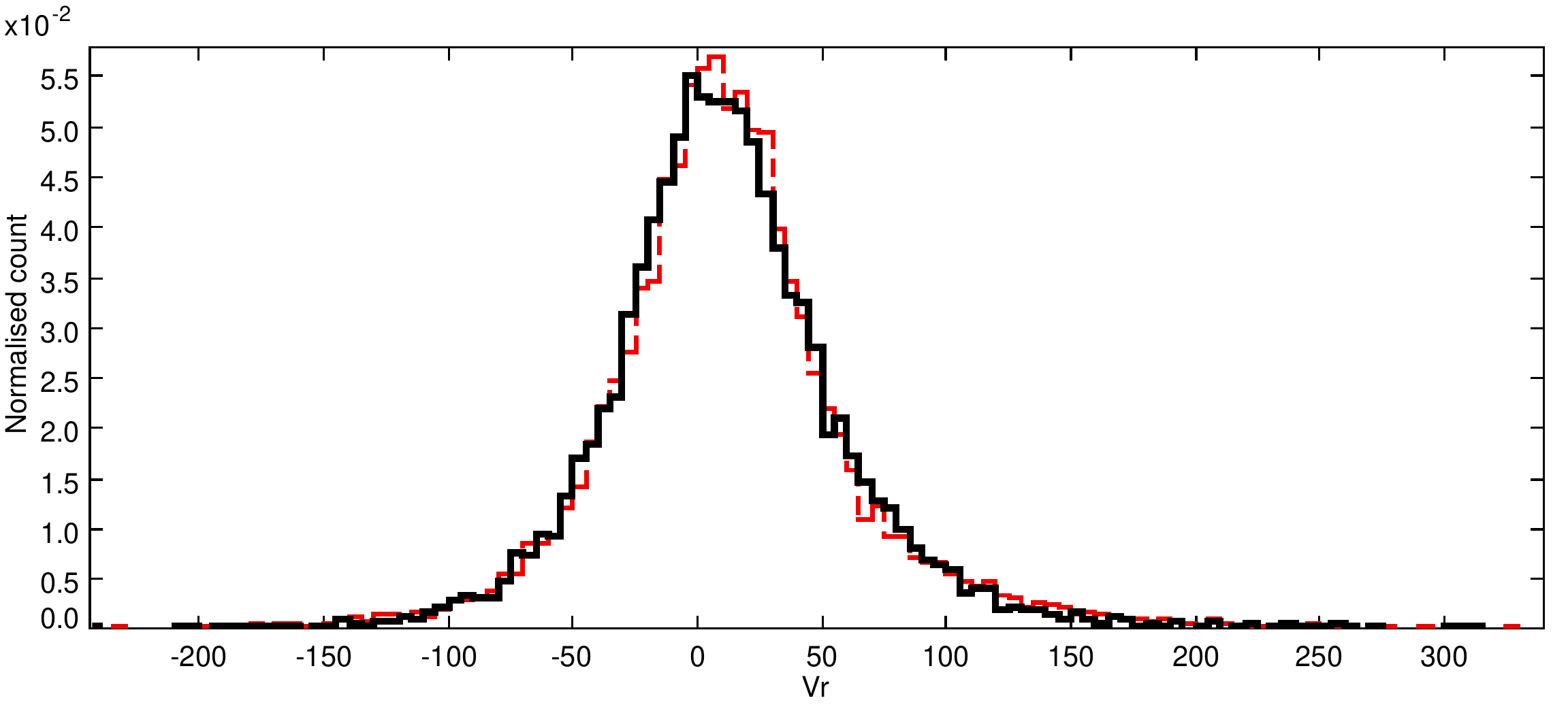}
\includegraphics[width=8cm,angle=0]{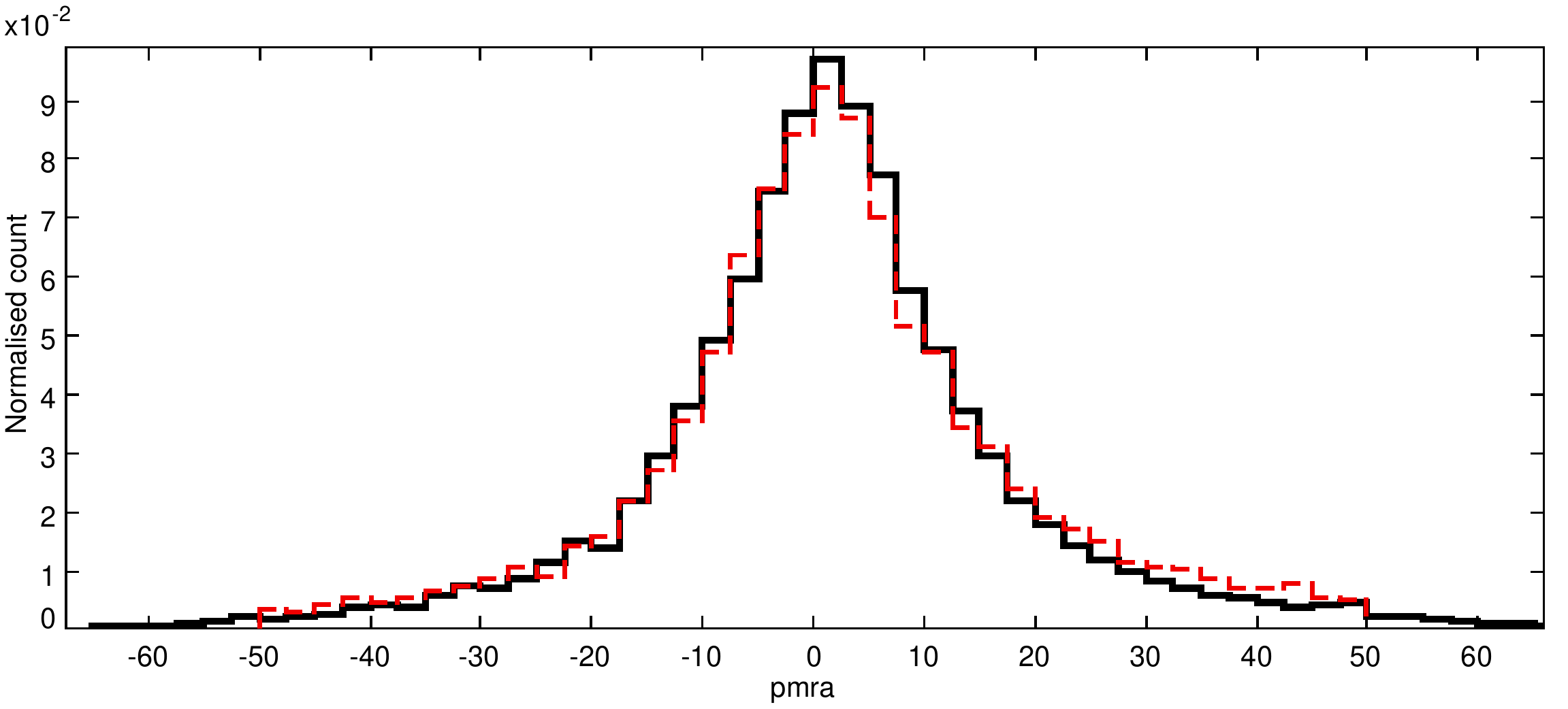}\includegraphics[width=8cm,angle=0]{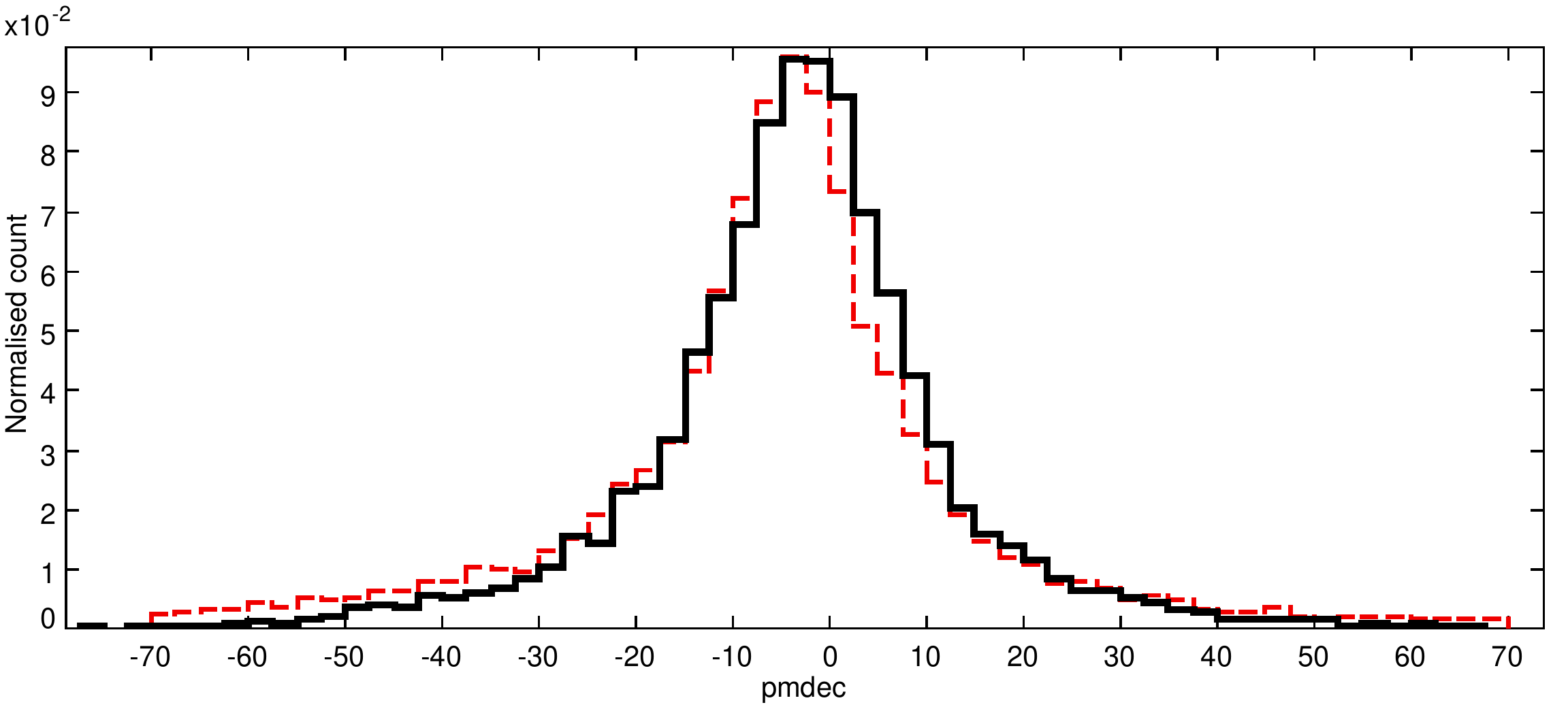}
\caption{{Same as Fig.~\ref{fig_feh-08}} for the metallicity bin -0.8 to -0.4 dex, which is dominated by the main thick disk. }
\label{fig_feh-04}
\end{center}
\end{figure*}

\begin{figure*}[htb]
\begin{center}
\includegraphics[width=8cm,angle=0]{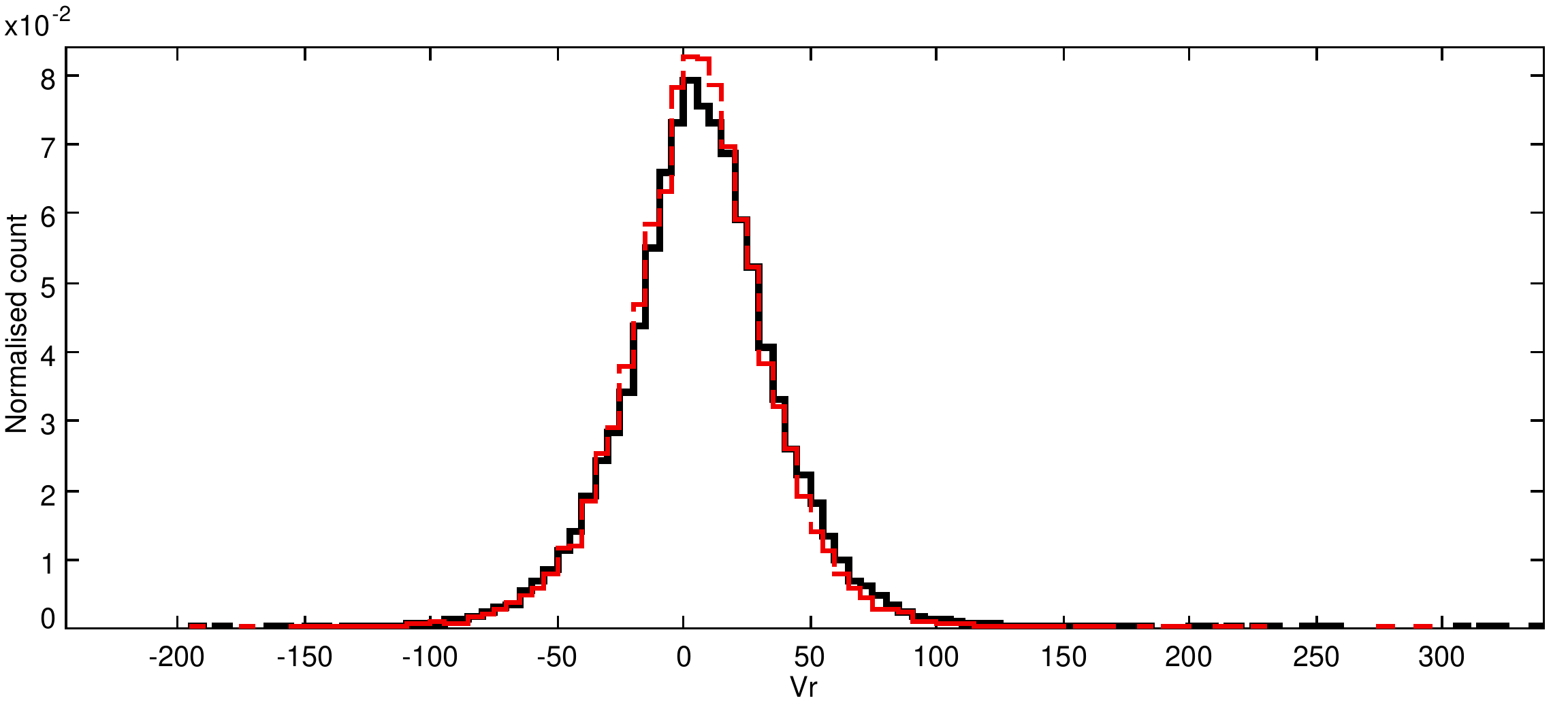}
\includegraphics[width=8cm,angle=0]{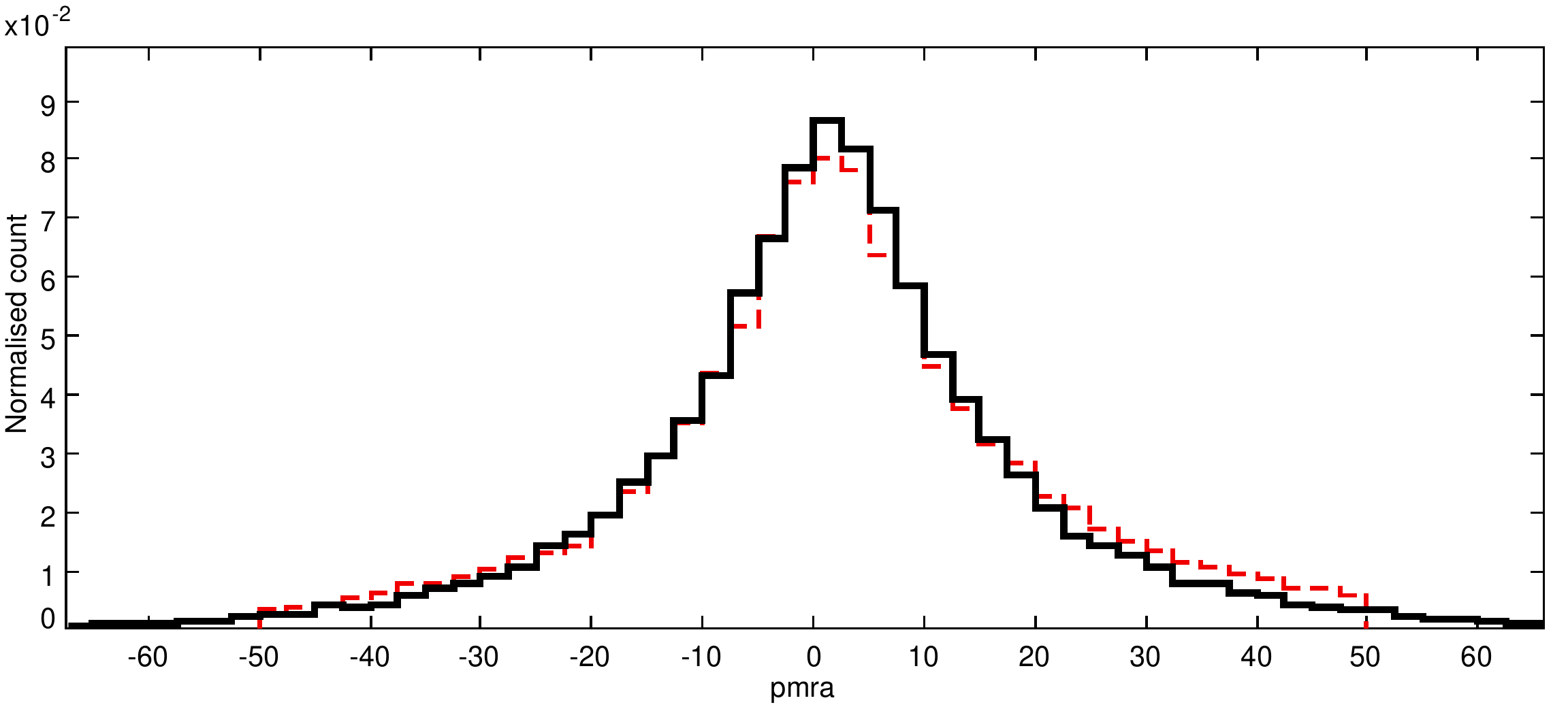}\includegraphics[width=8cm,angle=0]{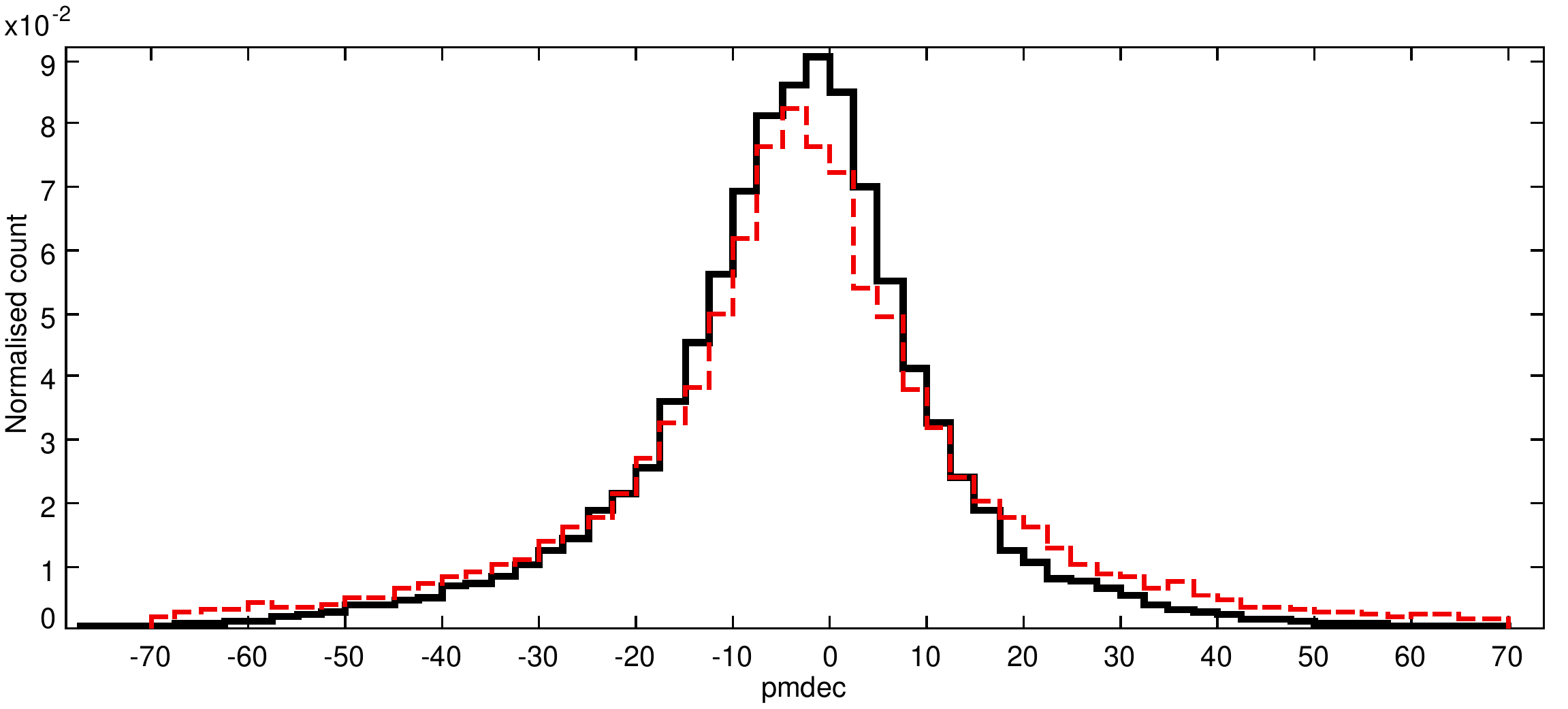}
\caption{{Same as Fig.~\ref{fig_feh-08}} for the metallicity bin -0.4 to 0 dex, which is dominated by the main thin disk.}
\label{fig_feh0}
\end{center}
\end{figure*}

\begin{figure*}[htb]
\begin{center}
\includegraphics[width=8cm,angle=0]{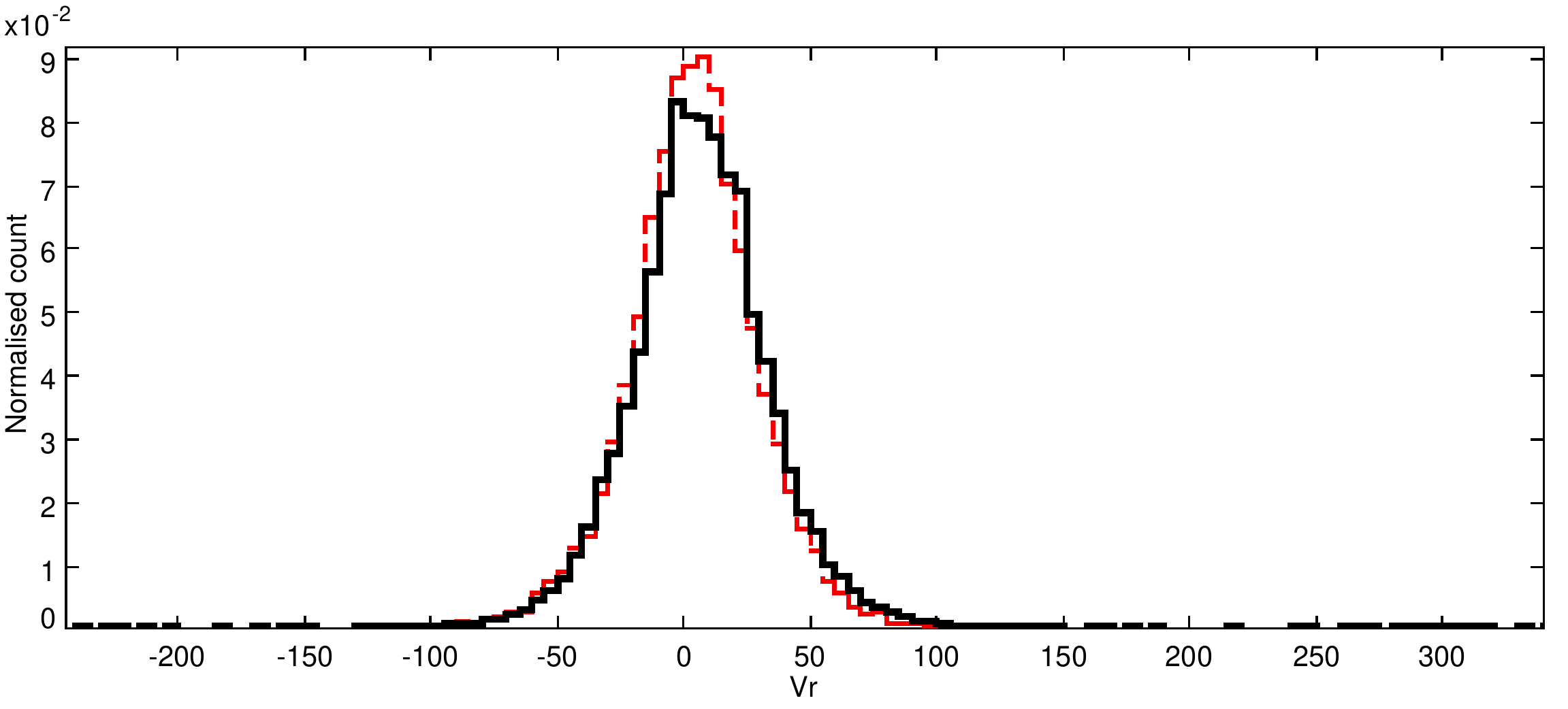}
\includegraphics[width=8cm,angle=0]{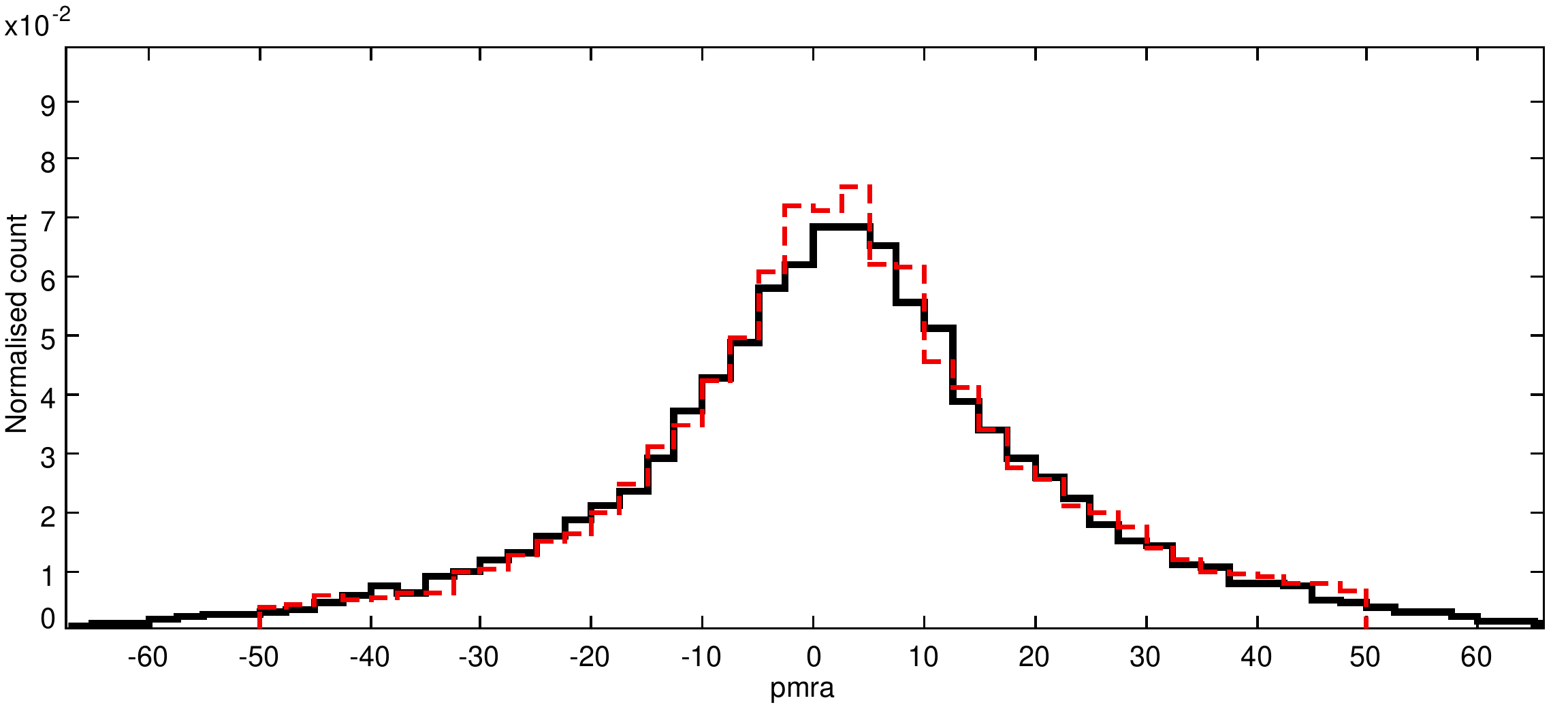}\includegraphics[width=8cm,angle=0]{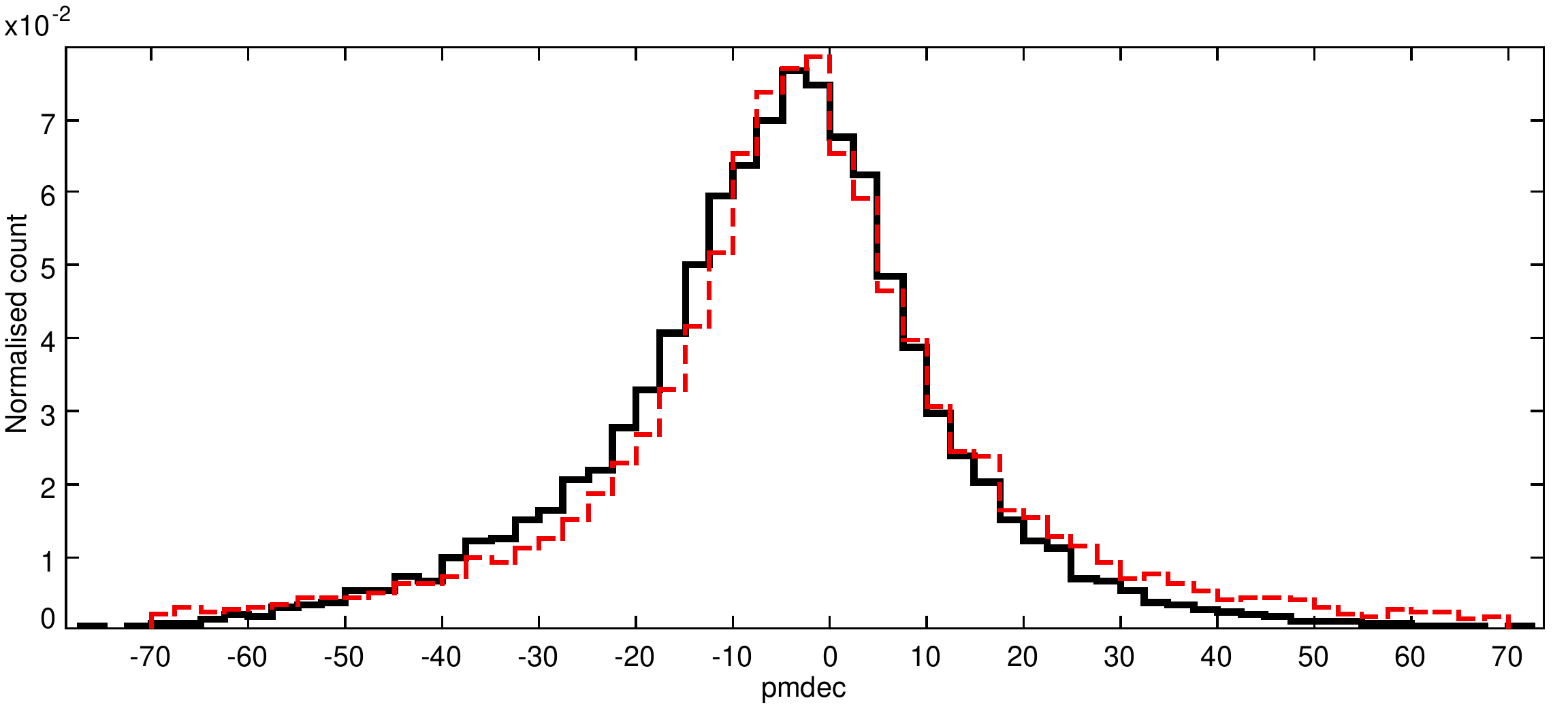}
\caption{{Same as Fig.~\ref{fig_feh-08}} for the metallicity bin 0 to 0.4 dex, which is dominated by the metal-rich thin disk. }
\label{fig_feh04}
\end{center}
\end{figure*}

Histograms of radial velocity and proper motions by metallicity bins are shown in Figure~\ref{fig_feh-08} for the metallicity range -1.2<[M/H]<-0.8 (dominated by the old thick disk), in Figure~\ref{fig_feh-04} for -0.8<[M/H]<-0.4 (dominated by the young thick disk), in Figure~\ref{fig_feh0} for -0.4<[M/H]<-0 (mainly thin disk), and in Figure~\ref{fig_feh04} for -0.4<[M/H]<0.4 (metal-rich thin disk). 
We note that the fit to each individual population is good, with a higher dispersion for the old thick disk than for the young thick disk, as expected and seen in the dispersion in Table~3. For metal-rich stars, the proper motion in declination shows
a slight shift, which could be due to the vertex deviation, which
is expected to be higher in these (in the mean) younger stars. We consider to implement a spiral arm model in the future to investigate and solve this problem.


In order to have a complete view of the agreement of the model in different regions of the sky, histograms of RAVE radial velocities and TGAS proper motions in the 11 sky regions used are presented in Appendix A, where the data are plotted as solid lines and the model as dashed lines. Figures~\ref{histo-RV-cool}, \ref{histo-pm1-cool}, and \ref{histo-pm2-cool} show histograms for cool stars, while Figures~\ref{histo-RV-hot}, \ref{histo-pm1-hot}, and \ref{histo-pm2-hot} present similar plots for hot stars. In each figure, the columns indicate the metallicity range we used, from ]-1.2;-0.8] on the left to [0;0.4] on the right.{ This shows that the old (metal-weak) thick disk dominates in the first column, the main thick disk in the second column, the main thin disk in the third column, and the metal-rich thin disk dominates in the fourth column.}

The histograms show very good agreements between the model and the data in nearly all cases. There is noticeable Poisson noise in the old thick disk (first column) in many cases, which explains why the uncertainties on the parameters of this population are sensitively larger than for the thin (third and fourth columns) and young thick disks (second column). The hotter stars show significantly smaller dispersions than cooler stars in proper motions, which is mainly because the giants (cooler) are at larger distances in the mean. We also note the strong skewness of the distributions in many cases, which explains the necessity to fit the whole histograms and not only the mean and standard deviation of each parameter. 

 Although the global fits are very good, there are a few noticeable differences in some fields that are probably due to some substructures, such as streams, associations, or clusters.  It is also possible that our vertex deviation does not represent the real one well and needs to be better modeled, for example, with a spiral arm model. The regions that present systematic deviations with the TGAS proper motions are the South Galactic cap (shift and dispersion in $\mu_U$ and $\mu_V$), although the radial velocity dispersion is well reproduced. Even though TGAS is much more homogeneous and well behaved than previous astrometric surveys, it is not completely free of systematics, as shown by \cite{Arenou2017A&A...599A..50A}, especially because of the scanning law and the limited number of observations included in this first release. Future Gaia releases will allow us to solve this problem. Toward the Galactic center at intermediate latitudes (-45<l<45, 25<b<40), there is also a significant disagreement in $\mu_l$ for hot stars. However, the fit is nearly perfect in $\mu_b$ for all types of stars, cool and hot, and any metallicities. Proper motion histograms of hot stars also agree very well in all directions, but for the stars that are at larger distances, it might be harder to distinguish significant deviations from an axisymmetric model. The radial velocities from RAVE are very well reproduced by the model, especially for hot stars at all metallicities and in all directions. For low-metallicity cool stars, the histograms are noisy because of the small number of stars, but the histograms for high-metallicity bins are well reproduced in all regions.

\section{Discussion}

Compared with previous RAVE analysis, we have used different hypotheses that are improvements and probably give more reliable results. First, we use an improved asymmetric drift that explicitly depends on \Rgal~and \zgal~and on age and is consistent with the Galactic potential. Second, we use the metallicity to help distinguish the thin from the thick disk, and we separately consider temperature bins dominated by dwarfs and giants, respectively. Third, we explicitly take the selection bias of the data into
account and compare the model simulations in the space of observables. Fourth, we explore the full parameter space and also determine the radial velocity dispersion gradients (or kinematical scale length). 

Using this method, we obtain reliable results for the various tested parameters that describe the kinematics of stellar populations, as discussed below. We separately discuss the $V$ component of the solar motion, which is related to the circular velocity of the LSR, to the rotation curve, and to the asymmetric drift.

\subsection{U and W component of the solar motion}

We obtained the following values for the solar motion $U$\Sun= 13.2$ \pm 1.3 $ km/s. and $W$\Sun=7.1 $\pm 0.2$ km/s.

For the $U$\Sun  velocity, our value agrees well with previous results (\cite{Aumer2009MNRAS.397.1286A}: 9.96$\pm$ 0.33, \cite{Schonrich2010}: 11.1$\pm$0.7, \cite{Coskunoglu2011MNRAS.412.1237C}: 8.5 $\pm$ 0.4 km/s, \cite{Pasetto2012A&A...547A..70P}: 9.87 $\pm$0.37 km/s, \cite{Karaali2014PASA...31...13K}: 10 km/s, \cite{Sharma2014ApJ...793...51S}: 11.45$\pm$0.1, \cite{Bobylev2015BaltA..24...51B}: 6.0, \cite{Bobylev2016AstL...42...90B}: 9.12). The error bars do not generally take into account the systematics that are due to the model that was used. {\cite{Bobylev2016AstL...42...90B} }noted a slight degeneracy between $U$ and $V$ solar motion that is due to the portion of sky covered by RAVE.

The vertical solar motion is more reliably determined in most of the studies, with consistent values on the order of 7$\pm$ 1 km/s. We confirm this value.

\subsection{Thin-disk velocity ellipsoid and gradients}

 \cite{Williams2013MNRAS.436..101W} restricted their analysis to red clump stars in their study of
the RAVE survey. They studied in particular the differences between north and south, looking for non-axisymmetry, which we did not include. Evidence of non-axisymmetries has previously been pointed out by \cite{Siebert2011MNRAS.412.2026S} in the analysis of RAVE third data release \citep{Siebert2011AJ....141..187S}. We dedicate such an analysis to a future paper. In their figure~8, \cite{Williams2013MNRAS.436..101W} found that the asymmetric drift varies with \zgal~ by about 40 km/s between \zgal=0 and \zgal=2 kpc, which is in good agreement with our model. The lag in $V_\phi$ at the \zgal=0 is approximately constant with $R_{gal}$,
which also agrees with our results. These studies did not investigate the dependency of the velocity ellipsoid on age, as we did.
 
 In their study of a {RAVE internal data release, intermediate between DR3 and DR4}, \cite{Pasetto2012A&A...547A..71P} performed a detailed analysis of the velocity ellipsoid and cross-terms of the velocity dispersion and their variation with \Rgal~and $z_{gal}$. Similarly to our study, they did not find any evidence of variations with \Rgal, most probably because the RAVE sample is not deep enough to reach regions where the sensitivity to \Rgal~can be significant. In this study, the mean velocity dispersions for the thin-disk component were found to be (26, 20, and 16) along the $U,V,W$ components, respectively, which is in good agreement with our values because we studied its variation with age, while they did not. Their values are close to the values for the old thin disk.

\cite{Sharma2014ApJ...793...51S} investigated the disk kinematics from GCS and RAVE data analysis with an MCMC method, but with considerable differences with our results in the assumptions and in the fitting method. They used parameters from an older stellar population model \citep{Robin2003} introduced in the Galaxia model \cite{Sharma2011ApJ...730....3S}. Similar density laws are assumed for the thin and thick disks, but we have shown in \cite{Robin2014} that the parameters of the thick disk in \cite{Robin2003} are not accurate, although the impact of this assumption on their analysis might be minor. They tested distributions in velocities to be Gaussian or Shu distributions, and used the asymmetric formula proposed by \cite{Bovy2012ApJ...759..131B}. On the other hand, they assumed that the asymmetric drift does not depend on distance to the plane, which is different from our study.  They also fixed the scale length of the disk to be 2.5 kpc for both the thin and thick disk. While we used a scale length of 2.53 kpc for the thin disk, our thick disk was split into an old component of scale length 3 kpc and a younger component of scale 2 kpc. In contrast, we decided not to use the GCS survey to constrain the populations. For the GCS survey it is more difficult to correctly simulate the selection function, which can introduce critical biases.The study of \cite{Sharma2014ApJ...793...51S}  resulted in similar solar velocities in $U$ and $W$ but in a slightly higher thin-disk maximum vertical velocity dispersion of ~25 km/s. Figure~\ref{disc-diffusion} shows how it compares with data from  \cite{Holmberg2009} and \cite{Gomez1997} and with
the model of \cite{Bovy2012ApJ...759..131B}. The maximum velocity
of \cite{Sharma2014ApJ...793...51S}  for the thin disk is noticeably higher than our result and those of \cite{Bovy2012ApJ...759..131B} and  \cite{Gomez1997}.

\subsection{Thick-disk velocity ellipsoid}

Our study points toward a $\sigma_z$ of 28 km/s for the young thick disk and 59 km/s for the old thick disk. This is in fair agreement with most previous studies, although they generally do not separate the thick disk in this way.

\cite{Guiglion2015A&A...583A..91G} analyzed the Gaia-ESO Survey in order to study the dependency of the velocity dispersions on metallicities and abundances. The correspondence of the $\alpha$ abundance ratio with age should in principal lead to estimating the age-velocity dispersion relation. However, this correspondence is only qualitative at present. They showed that the velocity dispersion increases with $\alpha$, with at high $\alpha$ a higher value for lower metallicities (\feh $\approx$ -0.80) than at typical thick disk metallicity (\feh $\approx$ -0.53). We find similar results, although our lowest metallicity bin has a velocity dispersion $\sigma_W$ of 59 km/s, while they have about 50 km/s. For the standard thick disk with mean \feh of -0.5 we have $\sigma_W$ of 27 km/s, while they have about 35 km/s according to their Figure 9. However, it is difficult to compare their results with ours, which extensively explores the model parameter space, and because the selection function is very different between Gaia ESO survey and RAVE-TGAS data set. In general terms, however,
both studies find a similar trend of  $\sigma_W$ with \feh for the high $\alpha$ population.

{In a RAVE internal data release,} \cite{Pasetto2012A&A...547A..70P} found a mean thick-disk asymmetric drift of 49 km/s, in agreement with our model, where the lag explicitly depends on $z_{gal}$. In Figure~\ref{vca} we see that in our model the thick-disk lag reaches this value at \zgal~about 1.5-2 kpc at the solar radius. 
In their study the velocity ellipsoid of the thick disk is (56, 46, and 35), which is in between the two thick-disk components in the present study. 

\subsection{Solar circular velocity, asymmetric drift, and rotation curve}

$V$\Sun in particular has been widely discussed in the literature. Values range between about 3 km/s and 26 km/s. Of the most recent studies, \cite{Binney2010MNRAS.401.2318B}  study based on GCS and  SDSS data points toward 11 km/s,  \cite{Schonrich2010} also from GCS toward 12.24$\pm$ 0.5 km/s, but with a sensitively different model. On the lower side limit, \cite{Aumer2009MNRAS.397.1286A} used Hipparcos data and found $V$\Sun 5.25$\pm$ 0.54 km/s, while \cite{Bovy2012ApJ...759..131B}, using preliminary APOGEE data (first year release), obtained 26$\pm3$ km/s, but with a high value of $V_{LSR}+V$\Sun=242 km/s. 

{Using a RAVE internal data release with 402 721 stars, close to the final DR4, \cite{Golubov2013A&A...557A..92G} found V\Sun=3 km/s using a new formulation of the Str\"omgren relation.  From the final RAVE-DR4, }
\cite{Karaali2014PASA...31...13K} analyzed the dependency of the determination of $V$\Sun on the Z coordinate and found a minimum of $\approx$ 10 km/s in the plane. {\cite{Wojno2016MNRAS.461.4246W} studied the chemical separation of disk components. They showed the dependency of the asymmetric drift (their Fig. 4ab) with [Fe/H],  drastically different for the thin and thick components defined according to their [Mg/Fe] abundances. {They assumed the solar motion from \cite{Schonrich2010}. Their result is difficult to compare with the present work. In our case the asymmetric drift is not a fitted parameter, and in contrast to \cite{Wojno2016MNRAS.461.4246W}, it  strongly depends on distance from the plane. Our values might also depend on metallicity via the age-vertical velocity dispersion relation, but this is not well known for the thick disk.}
 \cite{Jing2016MNRAS.463.3390J} used LAMOST data, which separate
the disk components based on the eccentricity of orbits. They obtained results that are similar to those of \cite{Wojno2016MNRAS.461.4246W}.}

\cite{Bobylev2015BaltA..24...51B} determined the solar motion from a local sample of young objects including masers and found $V$\Sun= 6.5 km/s. On the other hand, the same authors \cite{Bobylev2016AstL...42...90B} obtained quite significantly different values from the analysis of RAVE-DR4, based on Bottlinger's formulas and a determination of the distribution of stars in the $(U,V)$ plane using wavelet transform. $U$\Sun and $V$\Sun velocities were computed using UCAC4 proper motions and distances estimated by \cite{Binney2014MNRAS.437..351B}. They found $V$\Sun=20.8, but with $V$\Sun changing strongly with age and distance to the plane.

 \cite{Sharma2014ApJ...793...51S} obtained a range of values depending on whether they used Gaussian or Shu distributions, and on the degeneracies between parameters. An important point is that they did not easily distinguish the different populations, while we used metallicity and temperature to separate them.

 It should be noted that all investigations of the solar circular velocity are not made independently of assumptions or evaluations of the rotation curve and from the assumed asymmetric drift. Even the rotation curve at the position of the Sun is subject to debate. In studies based on local data it matters whether the rotation curve is flat, decreasing, or increasing at the Sun position. 
  
A preliminary analysis of the rotation curve from Gaia-TGAS data has been conducted by \cite{Bovy2017} from a completely different approach from the one that we undertook in our study. Their analysis did not solve for the solar motion and assumed the solar velocities from \cite{Schonrich2010}. They presented the Oort constant measurements and emphasized that the rotation curve slightly decreases at the solar Galactocentric radius. This is compatible with the rotation curves used in the present study. 
 
 Moreover, its determination using different tracers varies because
of neglected deviations from circular motion, or from neglected or unknown asymmetric drift variations with \Rgal~and $z_{gal}$. While it is well known and described for stars in the Galactic plane {\citep{Oort1965gast.conf..455O}, its variation with \zgal depends on the shape of the potential as well as on the cross terms of the velocity ellipsoids. With the notable exception of \cite{Binney2010MNRAS.401.2318B} and subsequent works \citep{Binney2012MNRAS.426.1328B,Binney2014MNRAS.439.1231B}, in other works the dependency of the asymmetric drift on $z$ is not included, neither empirically nor in a dynamically consistent way. This directly impacts these studies when they estimate of the velocity of the Sun relatively to the LSR.}
 In the present work, taking the strong variation of the asymmetric drift into account with \zgal~ dynamically self-consistently, we obtain a lower value for $V$\Sun than \cite{Schonrich2010} and several of the other studies cited above. We can explain the differences in these results as caused by the different stellar samples used in model comparisons or by the methods used. In many cases in the literature, some disagreements between models and data appear for the $V_\phi$ distributions (see for example \cite{Binney2014MNRAS.439.1231B}). The main difference in our case is the use of the observable space for model comparisons, while others are generally using distances, with their uncertainties and potential biases.

\subsection{Radial gradients of the velocity dispersion}

We attempted to constrain the radial gradient of $\sigma_U$ and $\sigma_W$ inside the thin-disk population, as they are ingredients of the asymmetric formula in the Galactic plane and can be related to the density scale length. However, we did not succeed to obtain this constrain because the standard deviation of our determination is far too large, probably because the ranges of Galactocentric distances available in RAVE and TGAS are too restricted to the solar vicinity. We will consider this point further when the Gaia DR2 are available.
 
\subsection{Vertex deviation}

{ Our fit included a vertex deviation, assuming two different values, for stars younger and older than 1 Gyr. 
Our result points toward a slight vertex deviation, but only at the two-sigma level, slightly higher for young stars than for older stars. This is not a strong signal that we do not discuss further. }

\section{Conclusions}
        
        In the approach that we present here, we used a full description of the asymmetric drift for each subcomponent of the BGM as a function of \Rgal~and \zgal, consistent with the velocity ellipsoid. This analysis shows that this lag is very important and varies strongly with \zgal~ and slightly with \Rgal. The comparison of the fitted histograms of radial velocity and proper motions shows that this method is efficient in reproducing the kinematics in a wide solar neighborhood. We draw the following conclusions: 
        
        \begin{itemize}
        \item The self-consistent description of the velocity ellipsoid based on an approximation of the St\"ackel potential provides a very efficient way to model the stellar kinematics in a wide region around the Sun. It reproduces the behaviors seen in RAVE and TGAS surveys very well.
        \item We provide new determinations of the solar motion, which confirms already known values of the velocity components $U$ and $W,$ and  propose a new value of the solar velocity toward rotation.
        \item We confirm the secular evolution of the velocity dispersion at the Sun found from Hipparcos data by \cite{Gomez1997}. 
        \item We do not obtain a strong constrain on the kinematics scale length. This requires further investigations using data at larger distances from the Sun. We are considering using APOGEE, GaiaESO, and future Gaia releases for this purpose.
        \item From these data alone, the constraint on the vertex deviation is not severe. A further investigation using the next Gaia releases is required.
        \end{itemize}
        
Hence, the combination of spectroscopic surveys providing accurate radial velocities (RAVE) with Gaia proper motions provide a strong revision of our view of the kinematics in the solar neighborhood. We expect that future Gaia data releases will enable us to extend this study to greater distances, to measure the kinematical gradients in particular, and to provide new constraints on the whole Galactic potential and to its approximation by St\"ackel potentials.

\begin{acknowledgements}

We are grateful to Francesca Figueras, Arnaud Siebert and Benoit Famaey for fruitful discussions. We acknowledge the financial support from "Programme National Cosmologie et Galaxies" (PNCG) of CNRS/INSU, France.
BGM simulations were executed on computers from the Utinam Institute of the
Universit\'e de Franche-Comt\'e, supported by the R\'egion de
Franche-Comt\'e and Institut des Sciences de l'Univers (INSU).

JGFT acknowledges support by the CNES (Centre National d'Etudes Spatiales) and by the R\'egion de Franche-Comt\'e.

Funding for RAVE has been provided by: the Australian Astronomical Observatory; the Leibniz-Institut fuer Astrophysik Potsdam (AIP); the Australian National University; the Australian Research Council; the French National Research Agency; the German Research Foundation (SPP 1177 and SFB 881); the European Research Council (ERC-StG 240271 Galactica); the Instituto Nazionale di Astrofisica at Padova; The Johns Hopkins University; the National Science Foundation of the USA (AST-0908326); the W. M. Keck foundation; the Macquarie University; the Netherlands Research School for Astronomy; the Natural Sciences and Engineering Research Council of Canada; the Slovenian Research Agency; the Swiss National Science Foundation; the Science \& Technology Facilities Council of the UK; Opticon; Strasbourg Observatory; and the Universities of Groningen, Heidelberg and Sydney.
The RAVE web site is at https://www.rave-survey.org.

This work has made use of data from the European Space Agency (ESA) mission Gaia (http: //www.cosmos.esa.int/gaia), processed by the Gaia Data Processing and Analysis Consortium (DPAC, http://www. cosmos.esa.int/web/gaia/dpac/consortium). Funding for the DPAC has been provided by national institutions, in particular the institutions participating in the Gaia Multi- lateral Agreement.

\end{acknowledgements}

\appendix
\section{Comparison of best model histograms with data}

We present histograms of radial velocities and proper motions in the 11 sky regions, comparing our best-fit model with real data. We separate stars by temperature, cool stars have $T_{\rm eff}$<5200K and hot stars $T_{\rm eff}$>=5200K.
In each figure, the columns correspond to different metallicity bins of width 0.4 dex from metal-weak thick disk [-1.2; -0.8] in the first column, ]-0.8; -0.4] dominated by the main thick disk in the second column, ]-0.4; 0] in majority the main thin disk in the third column, and [0; 0.4] the metal-rich thin disk in the fourth column. At the Galactic cap, the first and second components of the proper motions are parallel to $U$ and $V$ velocities, respectively. In other fields, the first and second components of the proper motions are the usual $\mu_l*$ and $\mu_b$ in mas/yr. We only present the results obtained with the first age-velocity dispersion (Eq. \ref{eq1}), as the other formulae give very similar results.

\begin{figure*}[htb]
\begin{center}
\includegraphics[width=24cm,angle=270]{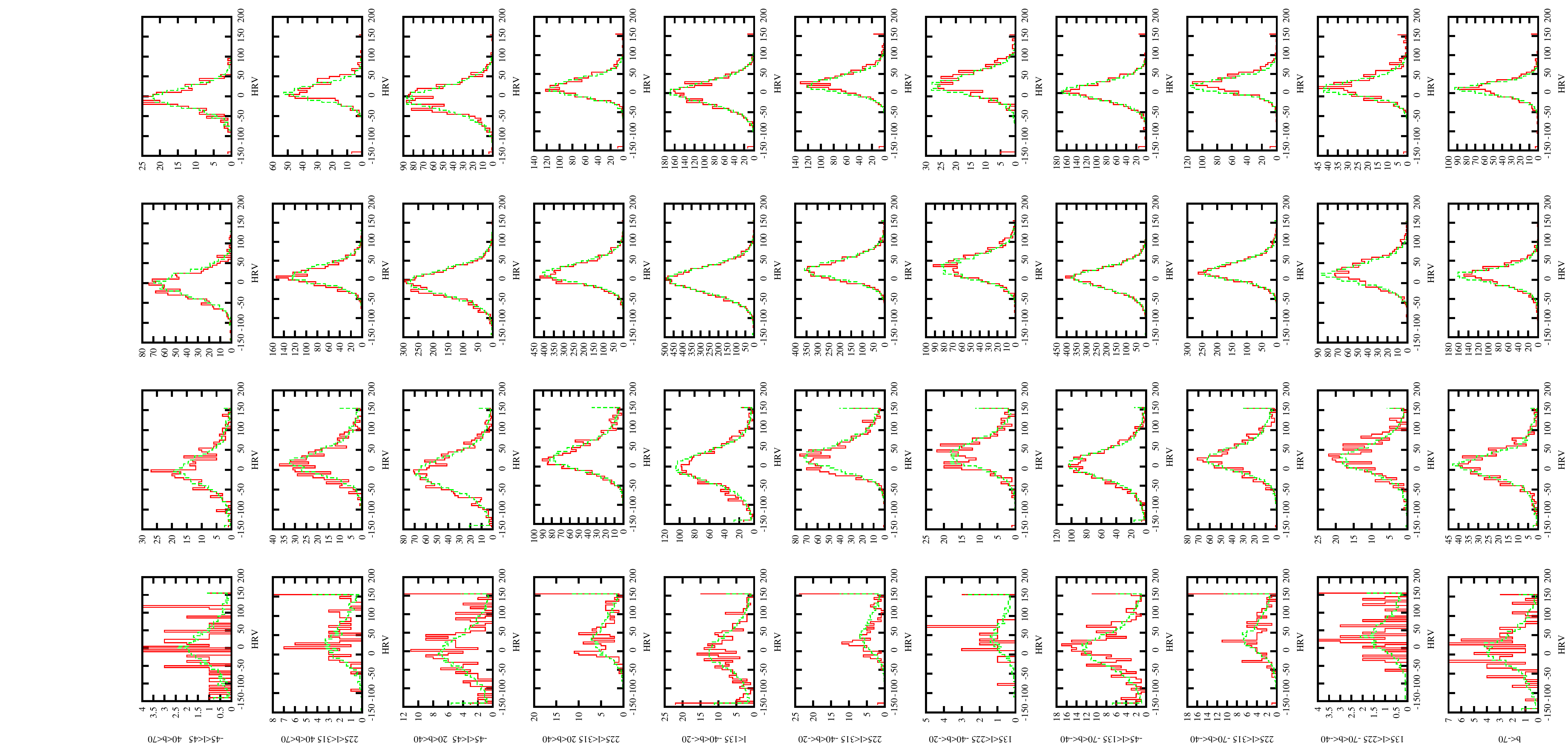}
\caption{Histograms of RAVE radial velocity distributions in different sky regions and for different metallicity bins, from -1.2 to 0.4 in steps of 0.4 dex for cool stars{, with a bin size
of 5 km/s}. Data are shown as red solid lines, and the best-fit model as green dashed lines. }
\label{histo-RV-cool}
\end{center}
\end{figure*}

\begin{figure*}[htb]
\begin{center}
\includegraphics[width=24cm,angle=270]{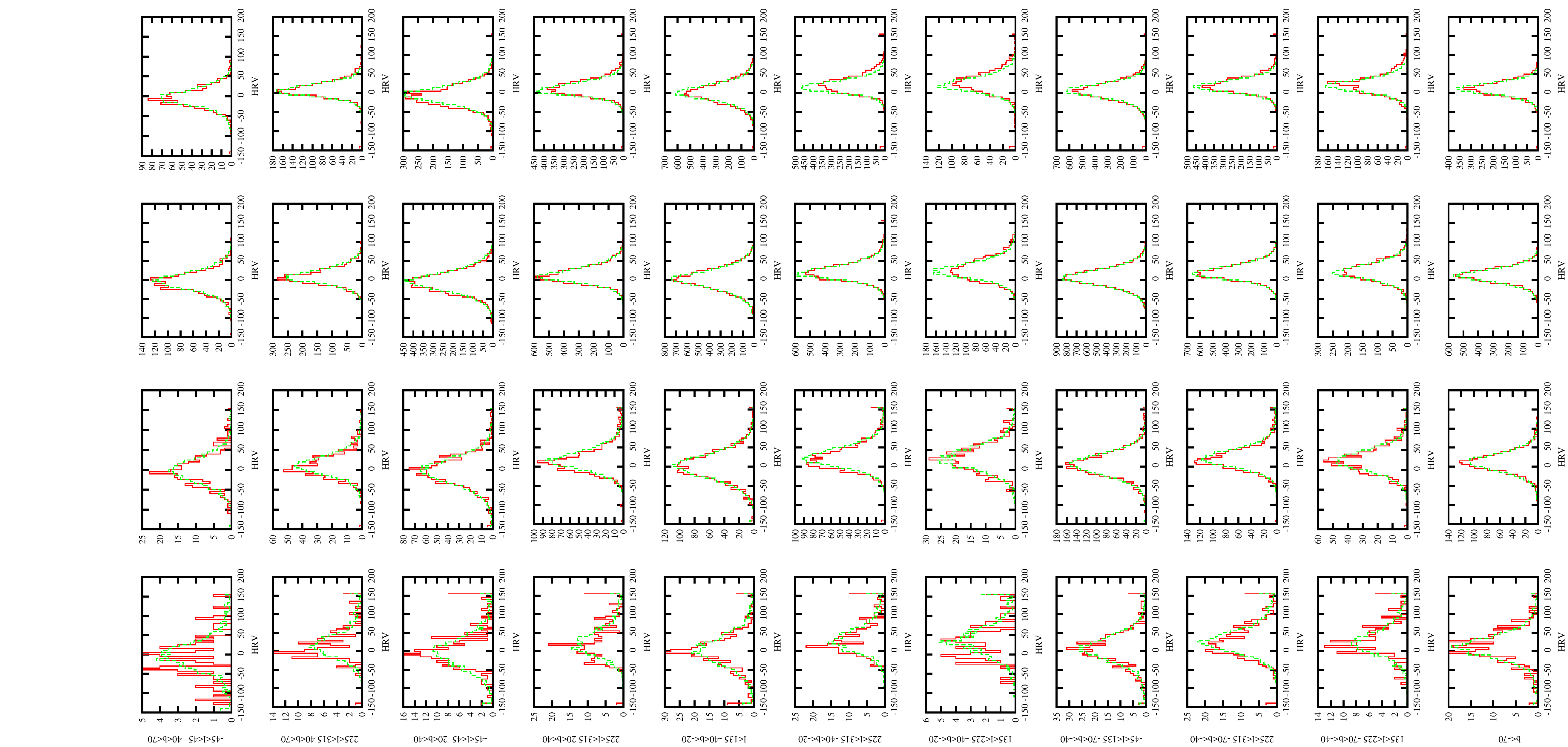}
\caption{Same as Fig. \ref{histo-RV-cool}, but for hot stars.}
\label{histo-RV-hot}
\end{center}
\end{figure*}
\begin{figure*}[htb]
\begin{center}
\includegraphics[width=24cm,angle=270]{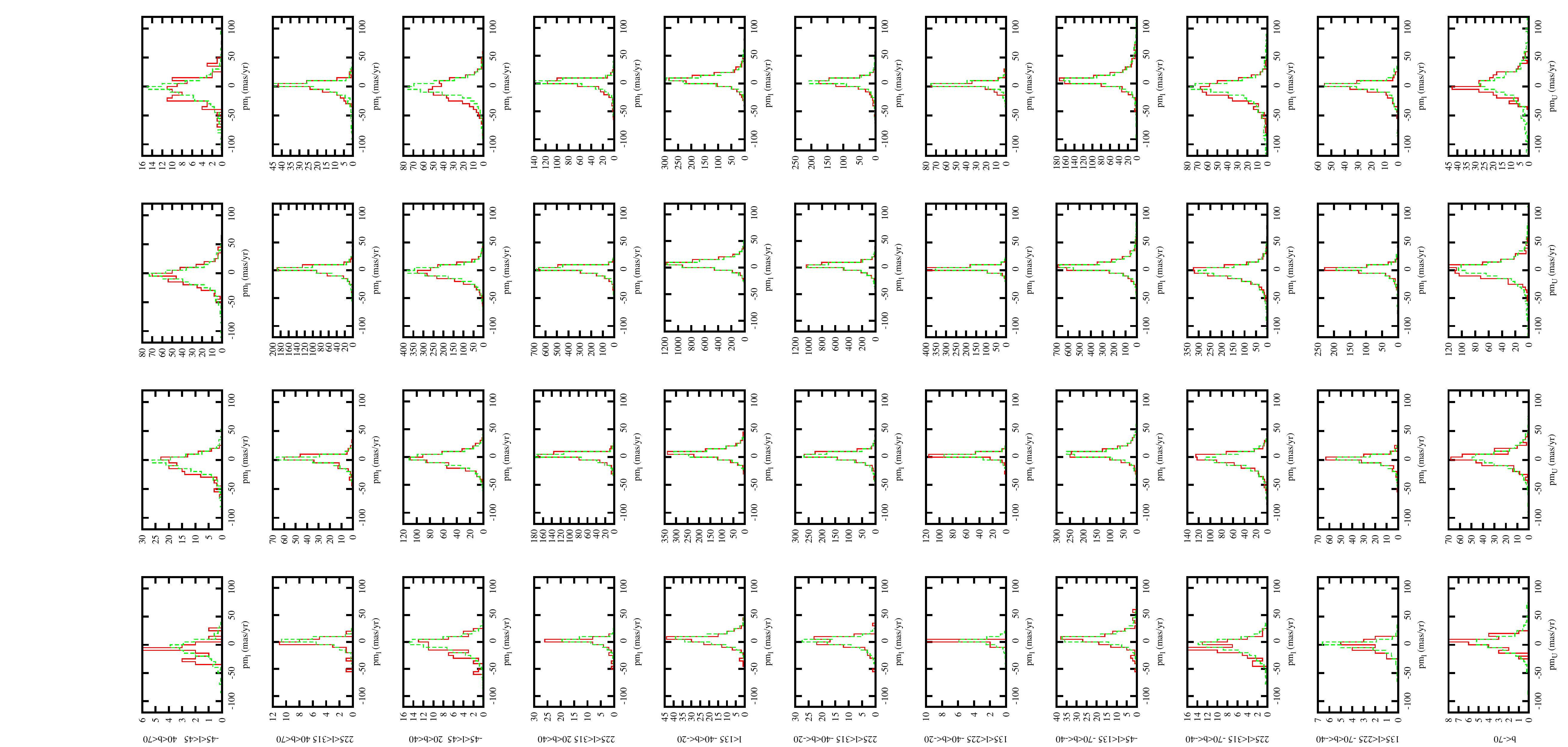}
\caption{Histograms of the first component of the TGAS proper motion distributions in different sky regions and for different metallicity bins (see Fig. \ref{histo-RV-cool}) for cool stars, {with a bin
size of 5 mas/yr}. Data are sown as red solid lines, and the best-fit model as green dashed lines. }
\label{histo-pm1-cool}
\end{center}
\end{figure*}
\begin{figure*}[htb]
\begin{center}
\includegraphics[width=24cm,angle=270]{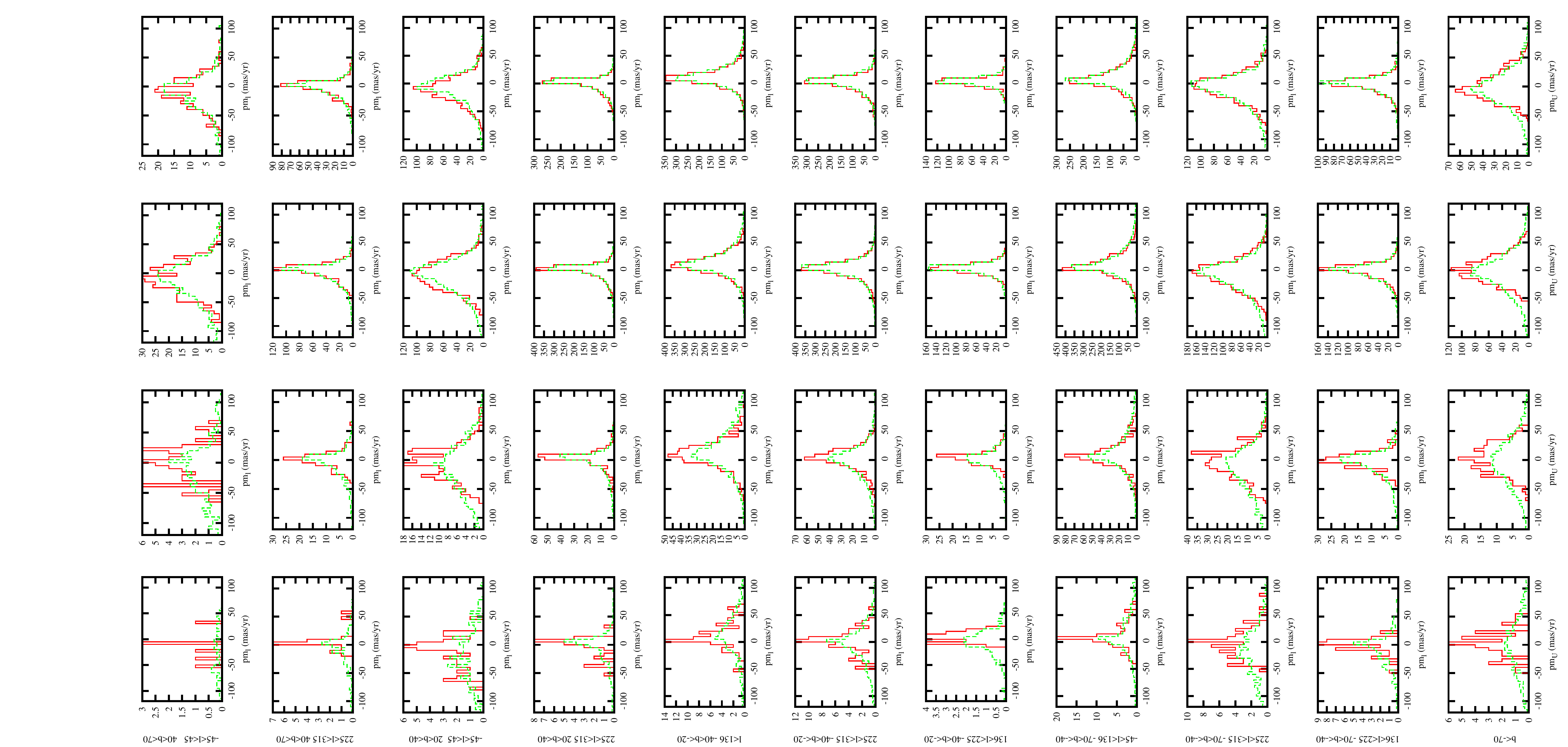}
\caption{Same as fig. \ref{histo-pm1-cool}, but for hot stars.  }
\label{histo-pm1-hot}
\end{center}
\end{figure*}
\begin{figure*}[htb]
\begin{center}
\includegraphics[width=24cm,angle=270]{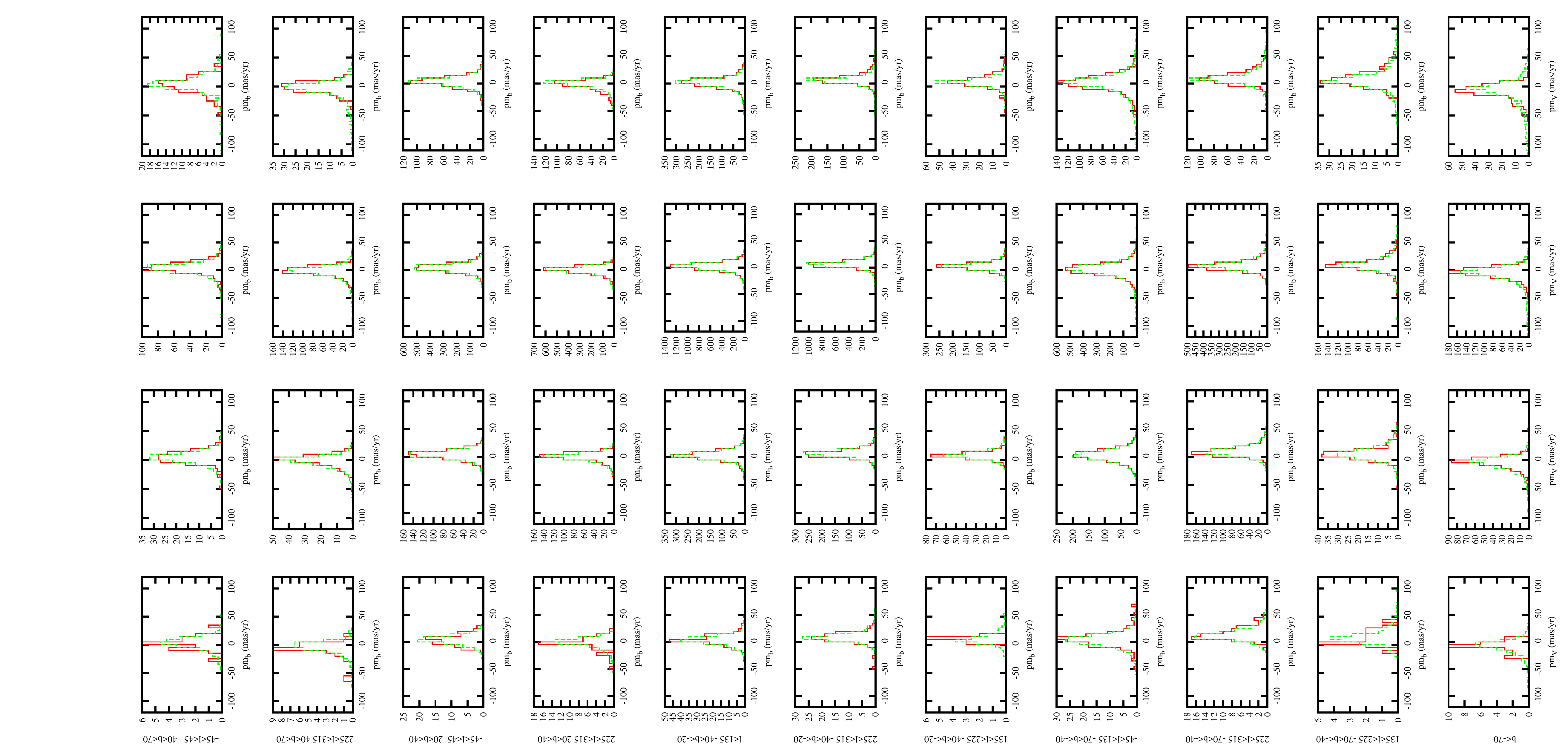}
\caption{Histograms of the second component of the TGAS proper motion distributions in different sky regions and for different metallicity bins (see Fig. \ref{histo-RV-cool}) for cool stars, {with a bin size of 5 mas/yr}. Data are shown as red solid lines,
and the best-fit model as green dashed lines. }
\label{histo-pm2-cool}
\end{center}
\end{figure*}\begin{figure*}[htb]
\begin{center}
\includegraphics[width=24cm,angle=270]{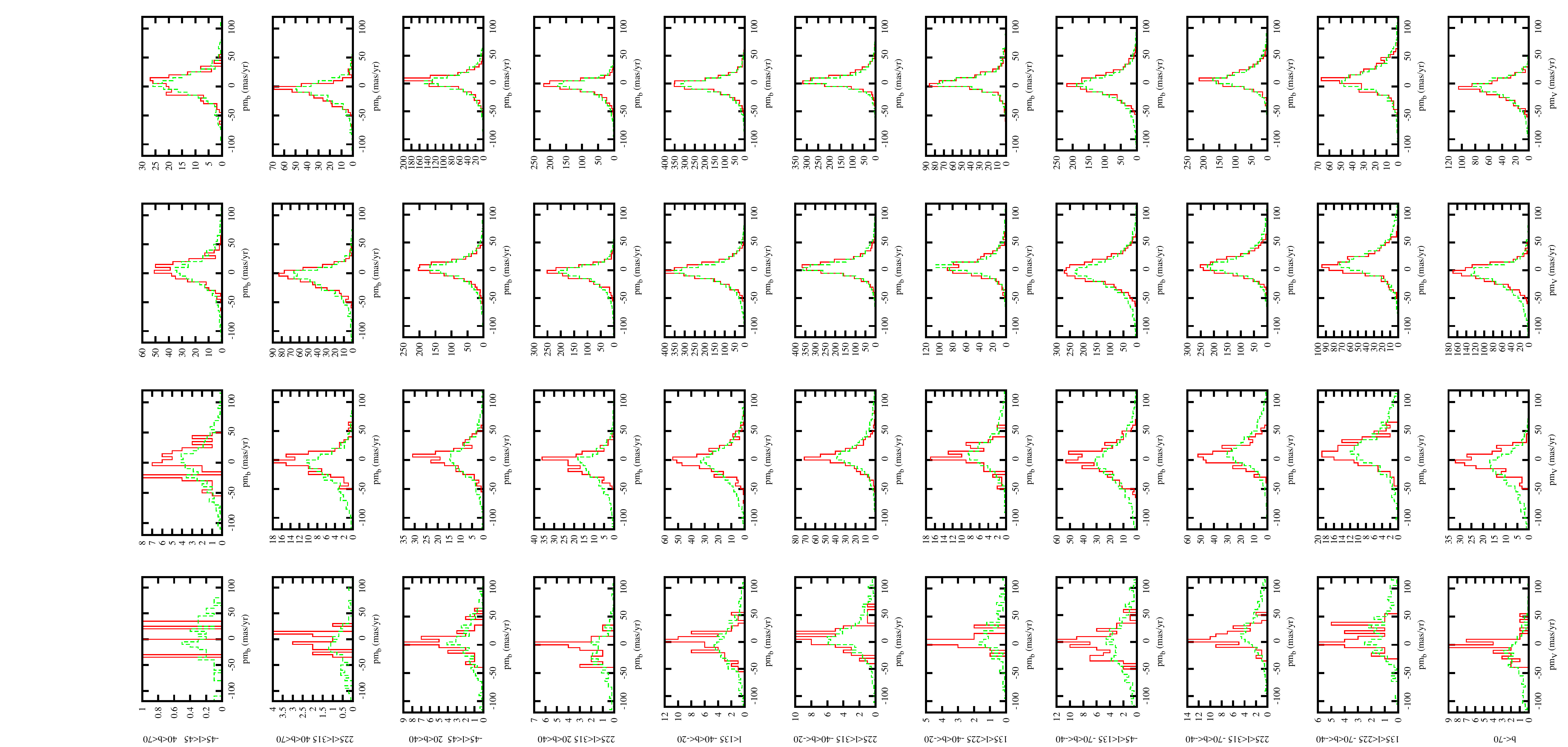}
\caption{Same as Fig. \ref{histo-pm2-cool}, but for hot stars.  }
\label{histo-pm2-hot}
\end{center}
\end{figure*}

\bibliographystyle{aa} 
\bibliography{RAVE-V2-astroph} 

\end{document}